\documentclass[superscriptaddress,groupedaddress,nofootnoteinbib,11pt]{article}

\usepackage{jcappub}
\usepackage{bm}
\usepackage{verbatim}

\usepackage{epsf}
\usepackage{graphicx,epsfig}
\usepackage{amsfonts}
\usepackage{amssymb}
\usepackage{xcolor}

\definecolor{mine}{rgb}{0.2,0.1,0.7}
\definecolor{bb}{rgb}{0.3, 0.5, 1}
\definecolor{bg}{rgb}{0.1, 0.1, 0.5}

\def\half{\frac12}

\def\L{\Lambda}

\def\L*{{\cal L}_*}
\def\L{\mathcal{L}}
\def\({\left(}
\def\){\right)}

\def\<{\langle}
\def\>{\rangle}

\newcommand{\bea}{\begin{eqnarray}}
\newcommand{\eea}{\end{eqnarray}}
\newcommand\be{\begin{equation}}
\newcommand\ee{\end{equation}}
\newcommand\beq{\begin{equation}}
\newcommand\eeq{\end{equation}}

\def\ba{\begin{eqnarray}}
\def\ea{\end{eqnarray}}

\newcommand{\refeq}[1]{(\ref{#1})}

\begin{document}

\title{On reaching the adiabatic limit in multi-field inflation}

\author[1,2]{S\'ebastien Renaux-Petel} 
\author[3]{and Krzysztof Turzy\'nski}
\affiliation[1]{Institut d'Astrophysique de Paris, UMR-7095 du CNRS, Universit\'e Pierre et Marie Curie, 98~bis~bd~Arago, 75014 Paris, France.}
 \affiliation[2]{Sorbonne Universit\'es, Institut Lagrange de Paris, 98 bis bd Arago, 75014 Paris, France}
  \affiliation[3]{Institute of Theoretical Physics, Faculty of Physics, University of Warsaw, Pasteura 5, \mbox{02-093 Warsaw}, Poland}

\emailAdd{renaux@iap.fr}
\emailAdd{krzysztof-jan.turzynski@fuw.edu.pl}

\keywords{inflation, cosmological perturbation theory.}

\vskip 8pt

\date{\today}


\abstract{We calculate the scalar spectral index $n_s$ and the tensor-to-scalar ratio $r$ in a class of recently proposed two-field no-scale inflationary models in supergravity. We show that, in order to obtain correct predictions, it is crucial to take into account the coupling between the curvature and the isocurvature perturbations induced by the noncanonical form of the kinetic terms. This coupling enhances the curvature perturbation and suppresses the resulting tensor-to-scalar ratio to the per mille level even for values of the slow-roll parameter $\epsilon \sim 0.01$. Beyond these particular models, we emphasise that multifield models of inflation are \textit{a priori} not predictive, unless one supplies a prescription for the post-inflationary era, or an adiabatic limit is reached before the end of inflation. We examine the conditions that enabled us to actually derive predictions in the models under study, by analysing the various contributions to the effective isocurvature mass in general two-field inflationary models. In particular, we point out a universal geometrical contribution that is important at the end of inflation, and which can be directly extracted from the inflationary Lagrangian, independently of a specific trajectory. Eventually, we point out that spectator fields can lead to oscillatory features in the time-dependent power spectra at the end of inflation. We demonstrate how these features can be model semi-analytically as well as the theoretical uncertainties they can entail.}

\maketitle

\section{Introduction}

Thanks to unprecedented accuracy of the measurements of cosmological
parameters \cite{Ade:2013uln,Komatsu:2014ioa}, it is now possible to begin to discriminate between
different models of cosmological inflation, whose predictions for the scalar spectral index $n_s$
must conform with the measured value
\begin{equation}
n_s = 0.960\pm0.007\, .
\label{eq:ns}
\end{equation}
The report of the observation of the $B$-mode of the cosmic microwave background by the BICEP2 team
\cite{Ade:2014xna}, although fully compatible with dust, as latter shown by the Planck collaboration \cite{Adam:2014bub}, has initiated a renewed effort in inflationary
model building, with the scope of accommodating the corresponding value of the tensor-to-scalar ratio $r=0.16^{+0.06}_{-0.05}$ in realistic setups.
In particular, there has been considerable activity 
along these lines, aimed at
extending 
the realisations of Starobinsky $R+R^2$ inflation
\cite{Starobinsky1,Starobinsky2,Starobinsky3}
in no-scale 
supergravity models
with a K\"ahler potential motivated by orbifold
compactifications of string theory.
Such extensions can utilise the
axionic component of the $T$ modulus
as the inflaton field, because this field
can have a quadratic potential after
all other degrees of freedom are 
stabilised \cite{Ferrara:2014ima,Kallosh:2014qta,Ellis:2014rxa, Hamaguchi:2014mza,Ellis:2014gxa,Ferrara:2014fqa}.

A model of this class, in which one field has a chaotic potential and the other has a Starobinsky-like potential, has been proposed in~Ref.~\cite{Ellis:2014gxa}. Here we use this model as a theoretically motivated and phenomenologically attractive starting point, and explore its original version,
as well as a number of slightly different variants in order to have a better understanding of the landscape of multi-field inflation. We
first calculate
the corresponding predictions for $n_s$ and $r$, using
the full set of equations of motion
for the inflationary perturbations. We show that in the absence 
of interactions specifically designed to stabilise the inflationary
trajectory and to give a large mass to the isocurvature perturbations, the value of $r$
predicted in the two-field model does not exceed the value
obtained in the original single-field $R+R^2$ Starobinsky model. The lesson for model-builders is clear: contrary to the initial hope of the authors of Ref.~\cite{Ellis:2014gxa}, simply `adding' fields with desirable inflationary potentials does not in general lead to predictions that interpolate between the individual single-field predictions. Even more important, it should be stressed that `adding' single-field inflationary models, each with well-defined theoretical predictions, is not sufficient to guarantee the existence of unambiguous predictions without additional theoretical inputs. The physical aspect at the origin of both phenomena is well known but is nonetheless frequently ignored in the literature: instantaneous isocurvature/entropic perturbations (we will use these two terms interchangeably) generically source the curvature perturbation (this feature was first pointed out in Ref.~\cite{Starobinsky:1994mh}). This implies that the curvature power spectrum evolves on super-Hubble scales during inflation and that its value at the end of inflation is generically different from its naive value evaluated at Hubble crossing. It also implies that if some isocurvature perturbation is still present by the end of inflation, the model in question is not predictive without specifying a scenario for reheating, \textit{i.e.} that the latter stage can alter the properties of the cosmological fluctuations (see \textit{e.g.} Refs.~\cite{Leung:2012ve,Leung:2013rza,Meyers:2013gua} for recent studies). 

For this reason, we examine in detail in section \ref{adiabatic limit} the conditions that enabled one to actually derive predictions in the model under consideration, \textit{i.e.} how an adiabatic limit, in which isocurvature perturbations are negligible, is actually reached (see Refs.~\cite{Meyers:2010rg,Meyers:2011mm,Elliston:2011dr,Seery:2012vj} for related studies in other setups). The decay of the entropic modes is directly related to their effective mass squared, and we study their various contributions, trying to draw general lessons. In section \ref{sec:oscillations}, we slightly modify the model at hand to observe more diverse multifield effects. In particular, we exhibit a model in which entropic modes remain ineffective during most part of inflation, to have an important observational effect only at its very end. This model displays interesting features, such as an oscillating (in time) power spectrum, which we are able to understand semi-analytically. We eventually give our conclusions in section \ref{sec:conclusion}.

\section{Analysis of a two-field no-scale supergravity model}

We briefly review the theoretical material about multifield inflation that we will use throughout the paper, following Refs.~\cite{GrootNibbelink:2001qt,Gordon:2000hv,Lalak:2007vi,Langlois:2008mn}, and then analyse in detail the model put forth in the first version of Ref.~\cite{Ellis:2014gxa}.

\subsection{Two-field inflation in a nutshell}

We consider a collection of scalar fields $\phi^I$, endowed with a metric $G_{IJ}(\phi^K)$ on field space, minimally coupled to gravity and interacting through a potential $V(\phi^I)$(we use units in which $M_P=1$):
\begin{eqnarray}
S= \int  d^4x \,\sqrt{-g} \left( \frac{R}{2}  -\frac{1}{2}G_{IJ } \nabla_{\mu} \phi^I \nabla^{\mu} \phi^J -V(\phi^I) \right)\,.
\label{S2a}
\end{eqnarray}
At the background level, the scalar fields are taken to be homogeneous and evolving in a spatially flat universe with Friedmann-Lema\^itre-Robertson-Walker metric
\begin{equation}
ds^2=-dt^2+a^2(t)d{\vec x}^2\,,
\end{equation}
where $a(t)$ denotes the scale factor. The corresponding equations of motion read
\begin{eqnarray}
 3H^2&=&\frac12 \dot \sigma^2+V\,,  \\
 \dot{H}&=&-\frac12\dot \sigma^2\,, \\
{\cal D}_t \dot \phi^I  +3H  \dot \phi^I &+&G^{IJ} V_{,J}=0\,, \label{eom-scalars}
\end{eqnarray}
where dots denote derivatives with respect to $t$, $H \equiv \dot a/a$ is the Hubble parameter, $\frac{1}{2}\dot \sigma^2 \equiv \frac{1}{2}G_{IJ} \dot \phi^I \dot \phi^J$ is the kinetic energy of the fields, and, here and in the following, ${\cal D}_t A^I \equiv \dot{A^I} + \Gamma^I_{JK} \dot \phi^J A^K$ for a field space vector $A^I$.\\

\noindent \textbf{Equations of motion in the natural field basis}.--- The dynamics of linear cosmological fluctuations about the above background is dictated by the second-order action \cite{Sasaki:1995aw,GrootNibbelink:2001qt,Langlois:2008mn}
 \begin{eqnarray}
S_{(2)}= \int  dt\, d^3x \,a^3\left(G_{IJ}\mathcal{D}_tQ^I\mathcal{D}_tQ^J-\frac{1}{a^2}G_{IJ}\partial_i Q^I \partial^i Q^J-M_{IJ}Q^IQ^J\right)\,,
\label{S2}
\end{eqnarray}
where the $Q^I$ are the field fluctuations in the spatially flat gauge and the mass (squared) matrix is given by
\begin{eqnarray} \label{masssquared}
M_{IJ} &=& V_{; IJ} - \mathcal{R}_{IKLJ}\dot \phi^K \dot \phi^L -\frac{1}{a^3}\mathcal{D}_t\left[\frac{a^3}{H} \dot \phi_
I \dot \phi_J\right]\,.
\end{eqnarray}
Here $ V_{;IJ} \equiv V_{,IJ}-\Gamma_{IJ}^K V_{,K}$, the Riemann tensor associated to the field space metric is denoted $\mathcal{R}_{IKLJ}$, and field space indices are raised and lowered using
$G_{IJ}$. One can easily deduce from equation \refeq{S2} the equations of motion for the linear fluctuations (in Fourier space):
\begin{eqnarray}
{\cal D}_t {\cal D}_t Q^I  +3H {\cal D}_t Q^I +\frac{k^2}{a^2} Q^I +M^I_{\,J} Q^J=0\,.
\label{pert}
\end{eqnarray}
The quantisation of the action \refeq{S2} and its practical numerical implementation to calculate cosmological power spectra has been exposed in various works (see \textit{e.g.} references \cite{Salopek:1988qh,GrootNibbelink:2001qt,Tsujikawa:2002qx,Weinberg:2008zzc,Huston:2011fr,McAllister:2012am,Amin:2014eta}) and we refer the interested reader to them for details. \\

\noindent \textbf{The adiabatic/entropic decomposition}.---Amongst the field fluctuations, the one pointing along the background trajectory in field space,
\begin{equation}
Q_{\sigma} \equiv e_{\sigma I} Q^I\,, \quad e_{\sigma}^I \equiv \dot \phi^I /{\dot \sigma}\,,
\end{equation}
is particular in that it is directly proportional to the comoving curvature perturbation $\zeta$,
\begin{eqnarray}
\zeta =\frac{H}{{\dot \sigma}} Q_{\sigma}\,.
\end{eqnarray}
The other fluctuating degrees of freedom, in the hyperplane perpendicular to the adiabatic direction (with respect to the field space metric), are referred to as entropic and embody the genuine multifield effects. In the following, we consider for simplicity two-field models, in which case there is only one entropic mode, and one can unambiguously define the entropic unit vector $e_s^I$ such that the orthonormal basis $(e_\sigma^I,e_s^I)$ is right-handed. It is straightforward to deduce from Eq.~\refeq{pert} the coupled equations of motion for $Q_\sigma$ and $Q_s \equiv e_{s I} Q^I$ (see \textit{e.g.} Refs.~\cite{Gordon:2000hv,Lalak:2007vi,Langlois:2008mn}). What will be of interest to us in the following are actually only their super-Hubble limit (\textit{i.e.} valid in the regime $k \ll a H$):   
\bea
&& \dot \zeta \simeq 2 \eta_{\perp} \frac{H^2}{\dot \sigma} Q_s \label{Rdot} \\
&&\ddot Q_s+3 H \dot Q_s +m^{2}_{s {\rm (eff)}} Q_s \simeq 0 \label{eqQs}
\eea
where we defined 
\be
\label{etaperpdefinition}
\eta_\perp  \equiv -\frac{V_{,s} }{H \dot \sigma}
\ee
and
\begin{equation}
\frac{m^{2}_{s {\rm (eff)}}}{H^2} \equiv \frac{V_{;ss}}{H^2} +3 \eta_\perp^2+ \epsilon  \, R^{{\rm field \, space}} \,.
\label{ms2}
\end{equation}
According to Eqs.~\refeq{Rdot}-\refeq{eqQs}, the entropic mode evolves on its own on super-Hubble scales, but it sources the curvature perturbation. The size of this coupling between the entropic and the adiabatic mode is measured by the dimensionless parameter $\eta_\perp$, which is non-zero whenever the inflationary trajectory deviates from a geodesic in field space (see \textit{e.g.} Refs.~\cite{Gordon:2000hv,GrootNibbelink:2001qt,Langlois:2008mn,Achucarro:2010da}). The evolution of the entropic mode itself is governed by the effective (super-Hubble) mass squared $m^{2}_{s {\rm (eff)}}$. There, $V_{;ss} \equiv e_s^I e_s^J V_{;IJ}$, $\epsilon \equiv -\dot H/H^2=\dot \sigma^2/(2 H^2)$ is the usual `deceleration' parameter and $R^{{\rm field \, space}}$ is the Ricci scalar of the field space metric $G_{IJ}$. Note eventually that in addition to the power spectrum $\cal P_{\zeta}$ of the curvature perturbation, we will also deal with the isocurvature power spectrum $\cal P_{{\rm iso}}$ of the rescaled variable $H/\dot \sigma \times Q_s$.

\subsection{Analysis of an exemplary model}

The model which is the starting point of the analysis
presented in Ref.~\cite{Ellis:2014gxa}
employs an effective two-field model described
by the Lagrangian:
\begin{equation}
\mathcal{L} = -\frac{1}{2}\partial_\mu\phi\partial^\mu\phi-\frac{e^{2b(\phi)}}{2}
\partial_\mu\chi\partial^\mu\chi - V(\phi,\chi)
\label{eq:lgen}
\end{equation}
with
$b(\phi)=\sqrt{2/3}\, \phi$
and
\begin{equation}
V(\phi,\chi) = \frac{3}{4}m^2\left( 1-e^{-\sqrt{2/3}\,\phi}\right)^2+\frac{1}{2}m^2\chi^2\, .
\label{eq:pot}
\end{equation}
In our numerical analysis, we employ the value $m=10^{-5}$ used
in Ref.~\cite{Ellis:2014gxa}, noting that the resulting power spectrum of
the curvature perturbations $\mathcal{P}_\zeta$ is proportional to $m^2$, so that
the normalisation of the perturbations can be adjusted to the observationally
determined value just by changing $m$. Hence the predictions of the model
for $n_s$ and $r$ depend only on the direction of the inflationary trajectory
giving a sufficient number of e-folds.

\begin{figure}
  \begin{center}
  \includegraphics*[width=11.5cm]{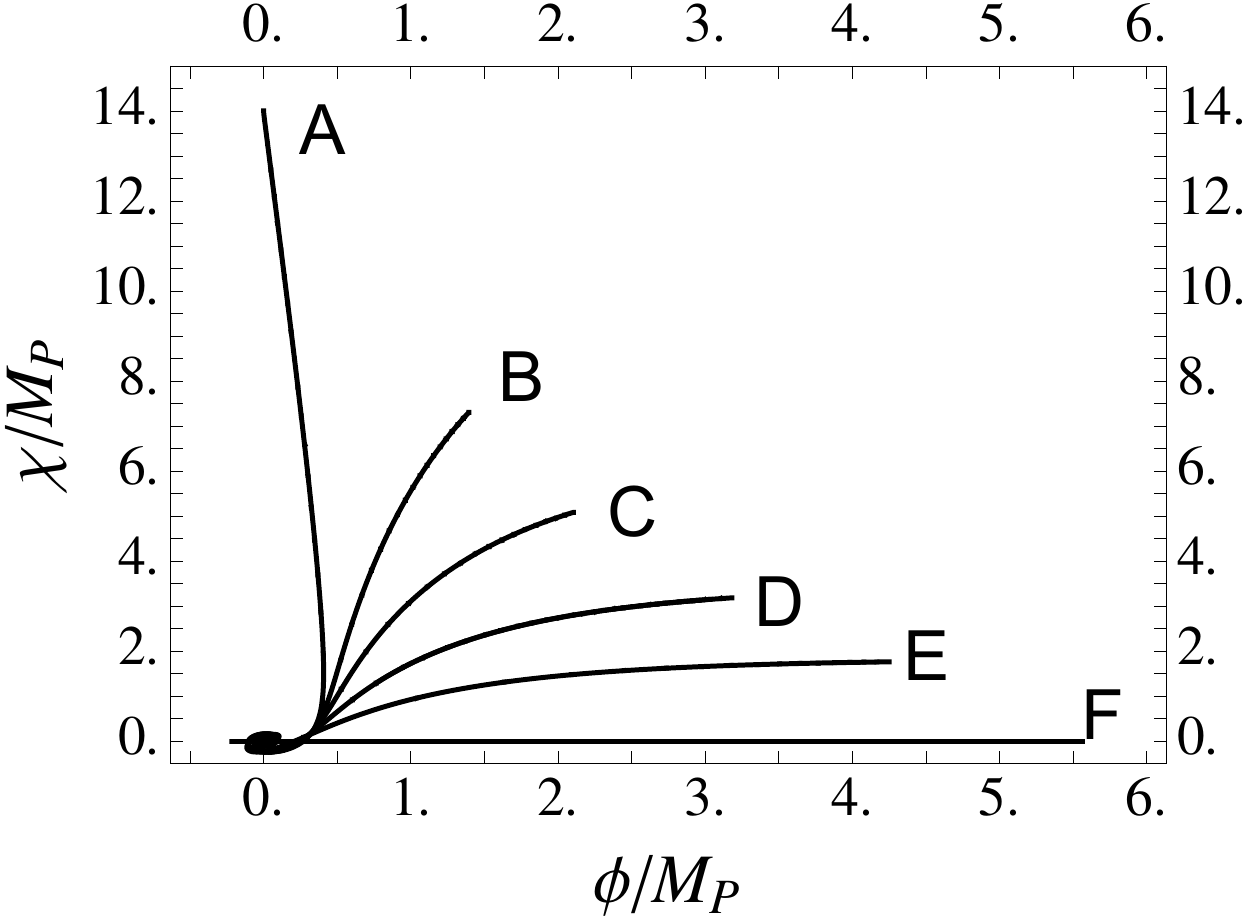}
\caption{A selection of inflationary trajectories in the field space giving more
than 60 e-folds. 
\label{f1}}
  \end{center}
\end{figure}

Examples of inflationary trajectories (labelled {\sf A} to {\sf F}) 
corresponding to different directions in field space are shown in Figure~\ref{f1} (we start with zero initial velocities).
For trajectory {\sf A} the field $\phi$ is initially zero (which is the minimum of its potential, {\em i.e.},
$\frac{\partial V}{\partial\phi}=0$ for $\phi=0$ and any value of $\chi$), but the interactions originating
from the non-trivial field space metric drive $\phi$ to nonzero values. Along trajectory {\sf F} one has $\chi=0$ and there is no coupling between the curvature
and the isocurvature perturbations, so this trajectory corresponds to the single-field
$R+R^2$ model.

\begin{figure}
 \begin{center}
\begin{tabular}{cc}
\includegraphics*[width=8cm]{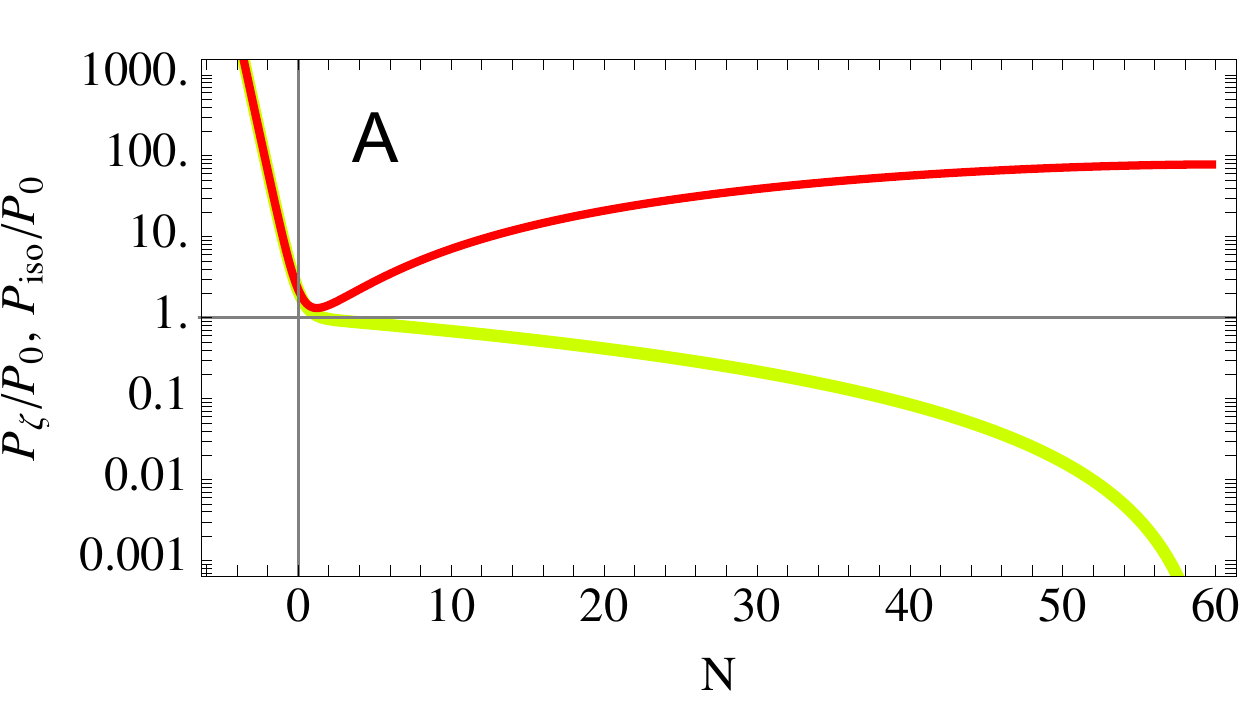}
&
\includegraphics*[width=8cm]{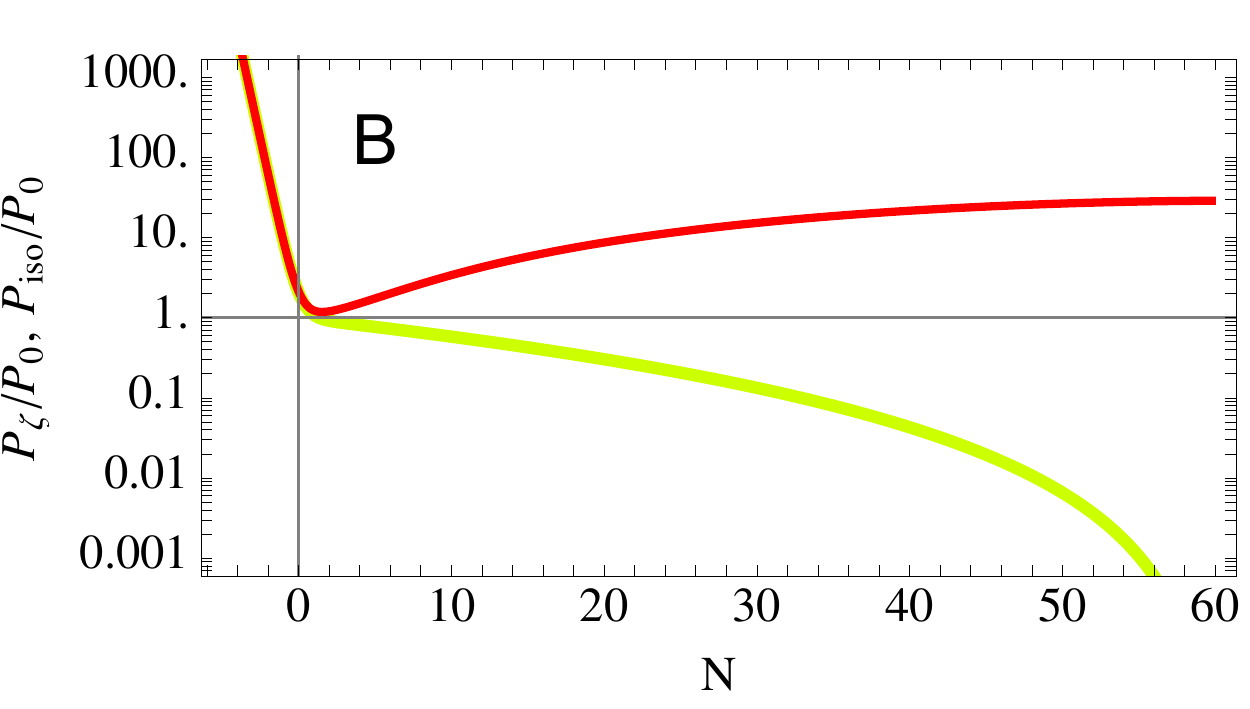}
\\
\includegraphics*[width=8cm]{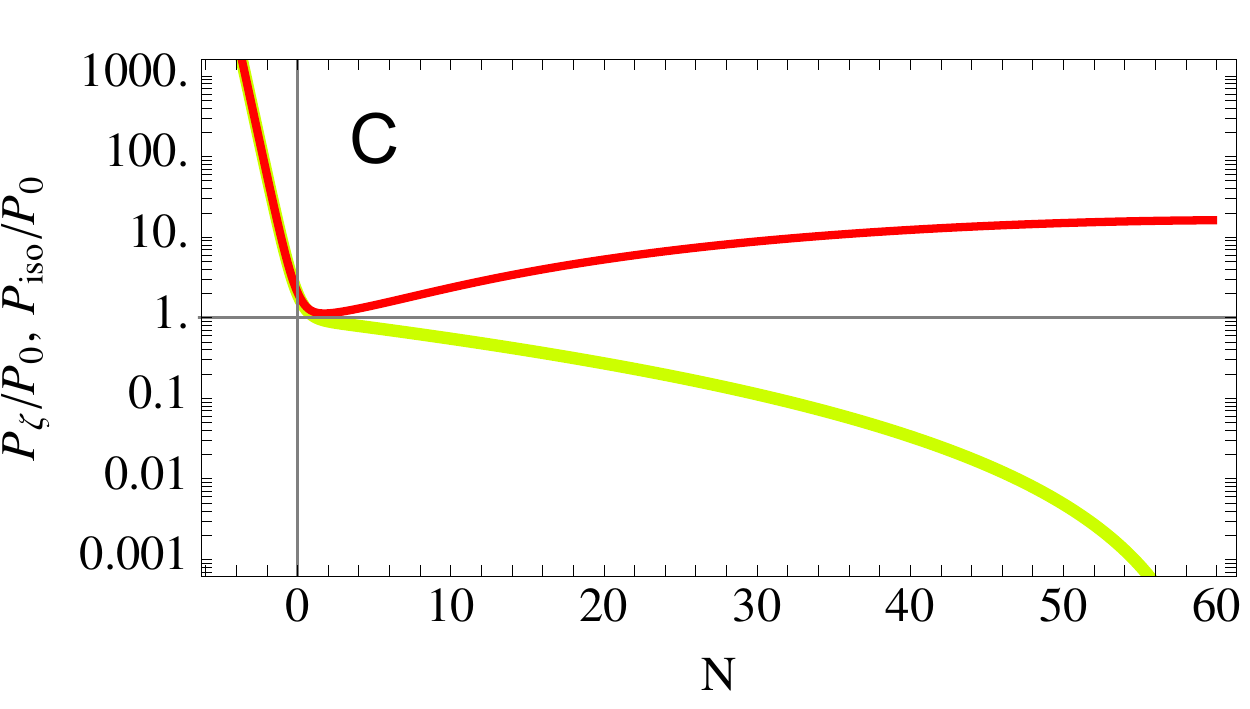}
&
\includegraphics*[width=8cm]{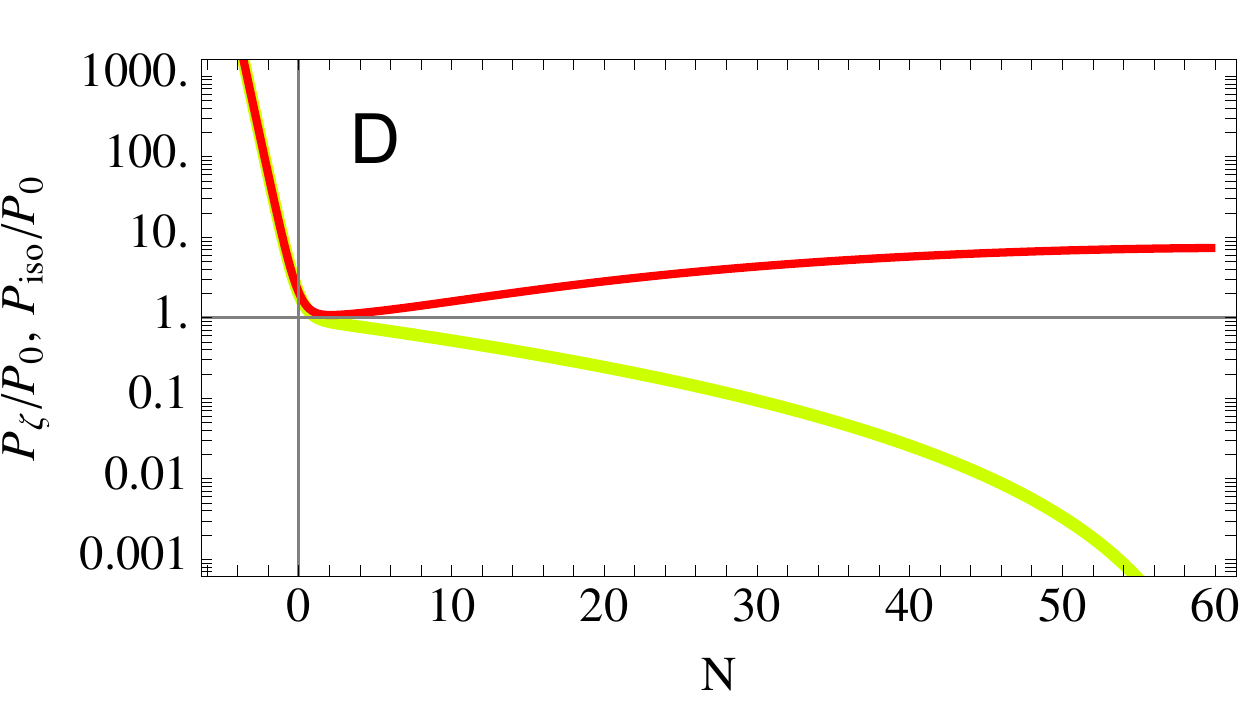}
\\
\includegraphics*[width=8cm]{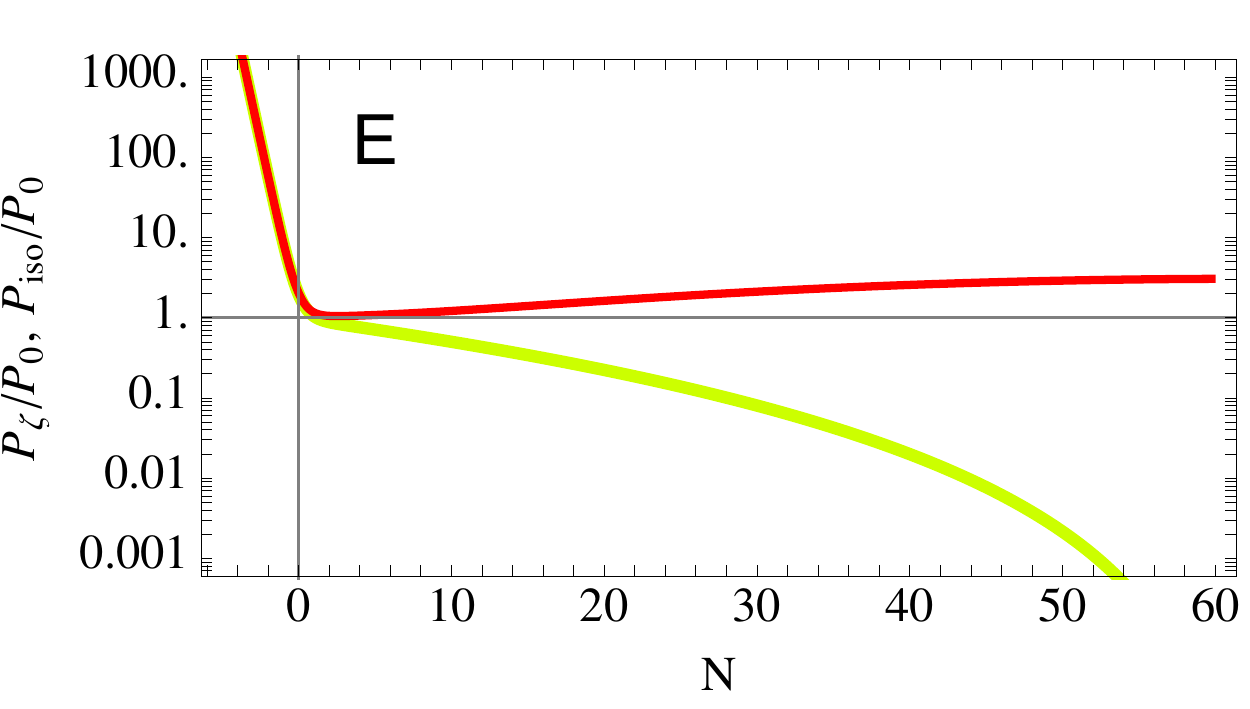}
&
\includegraphics*[width=8cm]{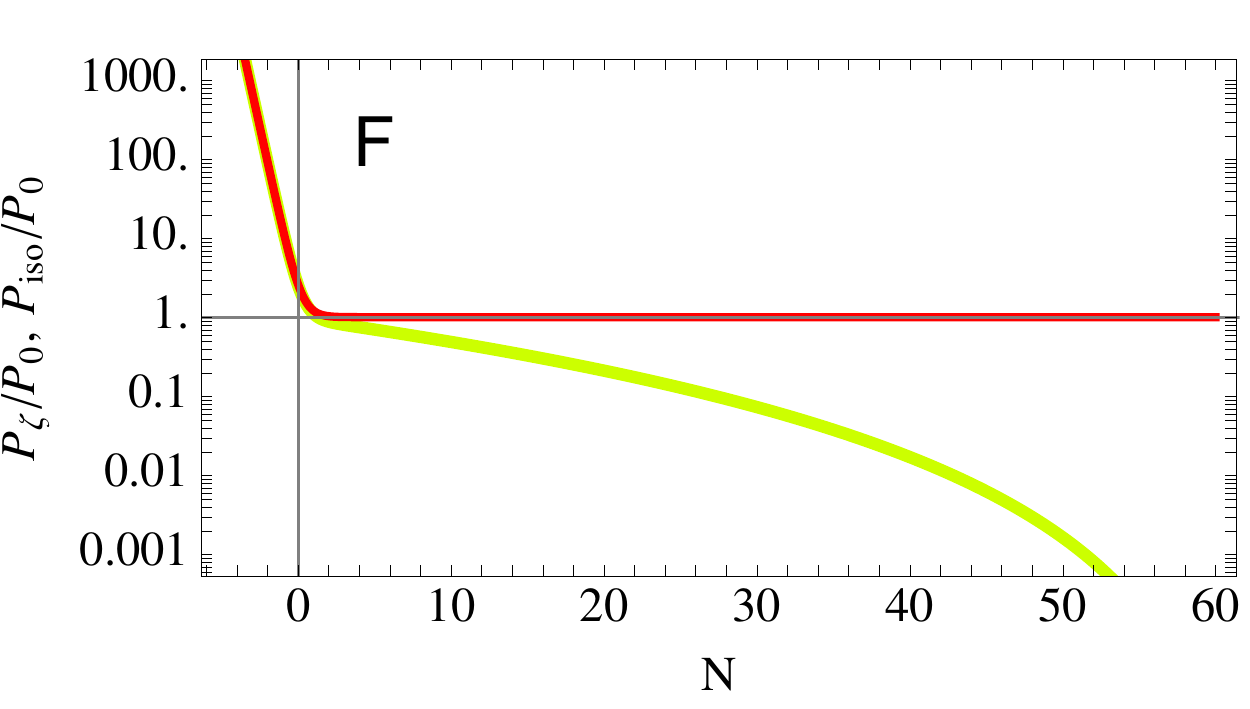}
\end{tabular}
\caption{For trajectories {\sf A} to {\sf F}: instantaneous power spectra of the curvature and isocurvature perturbations, $\mathcal{P}_\zeta$ (red lines) 
and $\mathcal{P}_\mathrm{iso}$ (yellow lines)
respectively, for the mode $k_{60}$ that leaves the Hubble radius 60 e-folds before the end of inflation (the latter is at $N=60$). The power spectra are given
in units of the single-field slow-roll prediction \mbox{$\mathcal{P}_0$}  given in Eq.~(\ref{eq:p0}) \label{f2}}
   \end{center}
\end{figure}

The evolutions of the instantaneous power spectra of the curvature perturbation, $\mathcal{P}_\zeta$,
and the isocurvature perturbation, $\mathcal{P}_\mathrm{iso}$, are shown in Figure~\ref{f2}
for the mode $k_{60}$ that leaves the Hubble radius 60 e-folds before the end of inflation, defined by $\epsilon=1$ (in this figure and in the followings, the time $N=60$ always denotes the end of inflation). The results are normalised
to the single-field slow-roll prediction:
\begin{equation}
\mathcal{P}_0 = \frac{H^2_\ast}{8\pi^2\epsilon_\ast M_P^2}\,,
\label{eq:p0}
\end{equation}  
where $\ast$ denotes evaluation at Hubble radius crossing.
It can be seen in Figure~\ref{f2} that the more the inflationary trajectory is directed along
the `noncanonical' direction $\chi$ in field space, the stronger is the sourcing of the curvature
perturbation by the isocurvature perturbation and the larger is the enhancement of the final
value of $\mathcal{P}_\zeta$. The enhancement occurs because the isocurvature perturbation around Hubble crossing is light compared to the Hubble scale, and the coupling between the perturbations,
proportional to $\eta_\perp$, can assume sizeable values for the inflaton
rolling down the `noncanonical' direction.
The values of the enhancement factors for each trajectory are given in Table~\ref{t1} for the modes $k_{50}$ and $k_{60}$ that crossed the Hubble radius respectively $N=50$ and $N=60$ e-folds before the end of inflation.

\begin{table}
\begin{center}
\begin{tabular}{|c|cccccc|}
\hline
final $\mathcal{P}_\zeta/\mathcal{P}_0$ & $\phantom{A}${\sf A}$\phantom{A}$ & $\phantom{A}${\sf B}$\phantom{A}$ & $\phantom{A}${\sf C}$\phantom{A}$ & $\phantom{A}${\sf D}$\phantom{A}$ & $\phantom{A}${\sf E}$\phantom{A}$ & $\phantom{A}${\sf F}$\phantom{A}$ \\
\hline
$k=k_{50}$ & 60 & 26 & 15 & 7.1 & 3.0 & 1.0  \\
$k=k_{60}$ & 74 & 28 & 16 & 7.3 & 3.1 & 1.0 \\
\hline
\end{tabular}
\end{center}
\caption{Enhancement of the final power spectrum of the curvature perturbation with
respect to the single-field slow-roll value $\mathcal{P}_0 $ given in (\ref{eq:p0})
for the modes $k_{50}$ and $k_{60}$ that crossed the Hubble radius respectively $N=50$ and $N=60$ folds before the end of inflation.  \label{t1}}
\end{table}

The enhancement of the curvature perturbation power spectrum  {\it via} sourcing by the
isocurvature perturbation is phenomenologically very relevant: if the final value of $\mathcal{P}_\zeta$ is dominated by the sourcing,
the resulting power spectrum inherits the statistical properties of the isocurvature perturbations
(see below a discussion for the case of the spectral index). The enhancement of $\mathcal{P}_\zeta$
also automatically results in a suppression of the tensor-to-scalar ratio~$r$, which now reads:
\begin{equation}
r = 16\epsilon_\ast\cdot\frac{\mathcal{P}_0}{\mathcal{P}_\zeta} \, .
\end{equation} 

\begin{figure}
  \begin{center}
  \includegraphics*[width=11.5cm]{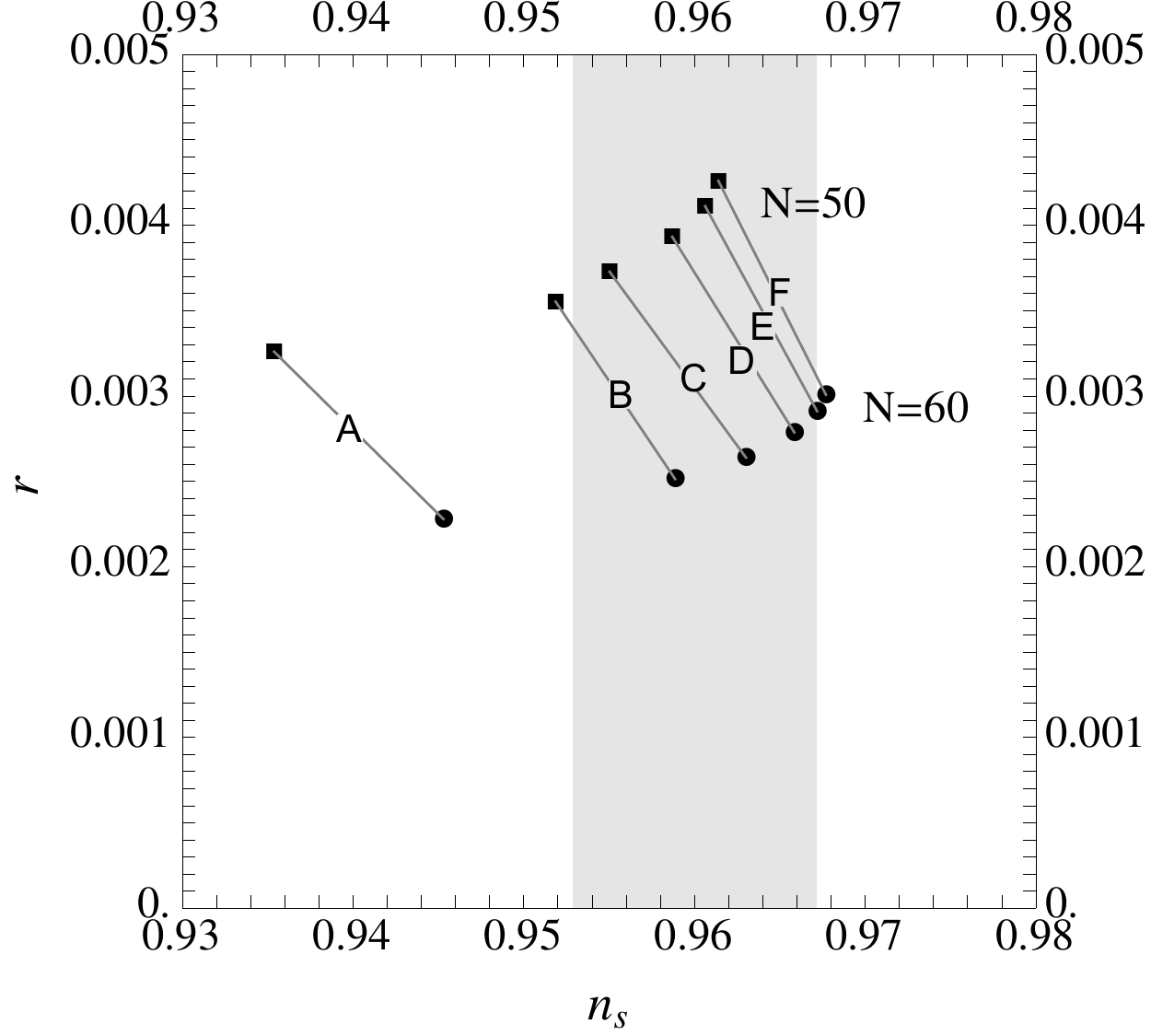}
\caption{Predictions for the scalar spectral index $n_s$ and the tensor-to-scalar ratio $r$ for the six 
inflationary trajectories shown in Figure~\ref{f1} with values of $N$ between 50~and~60. The shaded region
corresponds to the $1\sigma$ range of the Planck result for $n_s$ in Eq.~(\ref{eq:ns}). \label{f3}}
    \end{center}
\end{figure}

The predictions for $n_s$ and $r$ for the six inflationary trajectories {\sf A} to {\sf F} and for $N$ between 50~and~60 are shown in Figure~\ref{f3}. Although the values of $\epsilon_\ast$ range from 0.013 ({\sf A}, $N=50$) to $0.00018$ ({\sf F}, N=60), the predictions for $r$
are fairly uniform, at the level of a few per mille, due to the enhancement of the curvature perturbation by the isocurvature one. As a result,
none of the six trajectories are consistent with the BICEP2 measurement, when interpreted as being primarily due to primordial gravitational waves.\\

Besides the fully numerical calculations we made, it is interesting to note that accurate semi-analytical predictions for $n_s$ can be obtained. Using the large-scale equations of motion \refeq{Rdot}-\refeq{eqQs} and assuming a slow variation of the perturbations, one can indeed obtain the prediction \cite{Wands:2002bn,Langlois:2008qf}:
\be
n_s-1=-2\epsilon_*- \eta_* -2 \left(\eta_{\perp}\right)_* {\rm sin}(2 \Theta)+\left(\eta+\frac{2\, m^{2}_{s {\rm (eff)}}}{3 H^2} \right)_*{\rm sin^2} \Theta
\label{semi-analytical}
\ee
where we remind the reader that $*$ denotes evaluation at Hubble crossing $k=aH$, we have defined $\eta \equiv \frac{\dot \epsilon}{H \epsilon}$, and the angle $\Theta$ is defined such that 
\be
\frac{\mathcal{P}_0}{\mathcal{P}_\zeta}=\cos^2( \Theta)\,,
\ee
\textit{i.e} $\Theta=0$ if there is no super-Hubble transfer from entropic to adiabatic fluctuations, and $\Theta=\pi/2$ if the final curvature perturbation is mostly of entropic origin. We call this a semi-analytical prediction as it requires the knowledge of $\Theta$, \textit{i.e.} of the curvature perturbation at the pivot scale, which we calculate numerically. A comparison between the fully numerically calculated $n_s$ and the corresponding semi-analytical predictions can be found in Table~\ref{t2}: the agreement is excellent with errors less than $0.1\%$.\\

Before closing this section, let us note that, after the first version of this paper was online, the authors of Ref.~\cite{Ellis:2014gxa} reanalysed their model by taking into account the impact of entropic fluctuations. Their updated analysis in Ref.~\cite{Ellis:2014opa} agrees with ours.

\begin{table}
\begin{center}
\begin{tabular}{|c|ccccc|}
\hline
 & $\phantom{A}${\sf A}$\phantom{A}$ & $\phantom{A}${\sf B}$\phantom{A}$ & $\phantom{A}${\sf C}$\phantom{A}$ & $\phantom{A}${\sf D}$\phantom{A}$ & $\phantom{A}${\sf E}$\phantom{A}$ \\
\hline
$(n_s)_{\rm numerical}$ & 0.946 & 0.9591 & 0.962 & 0.9659 & 0.9672   \\
$(n_s)_{\rm semi-analytical}$ & 0.947 & 0.9598 & 0.963 & 0.9663  & 0.9676  \\
\hline
\end{tabular}
\end{center}
\caption{Comparison between the fully numerically calculated $n_s$ and the semi-analytical predictions Eq.~\refeq{semi-analytical} for trajectories {\sf A} to {\sf E} (the two values coincide for trajectory {\sf F}), at the pivot scale $k_{60}$.  \label{t2}}
\end{table}

\section{Reaching the adiabatic limit}
\label{adiabatic limit}

\begin{figure}[!h]
\begin{center}
\begin{tabular}{cc}
\includegraphics*[width=8.7cm]{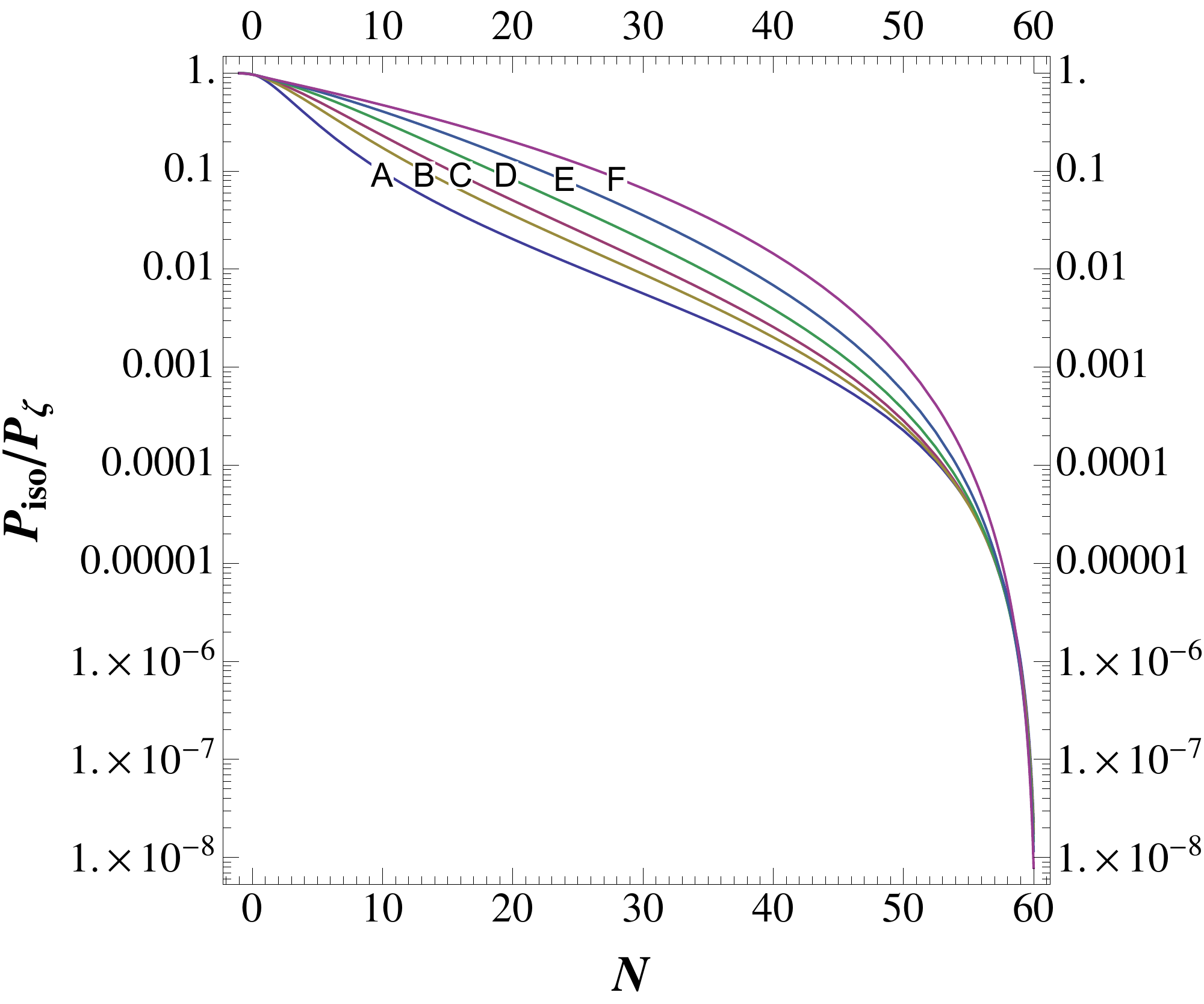}
&
\includegraphics*[width=7.5cm]{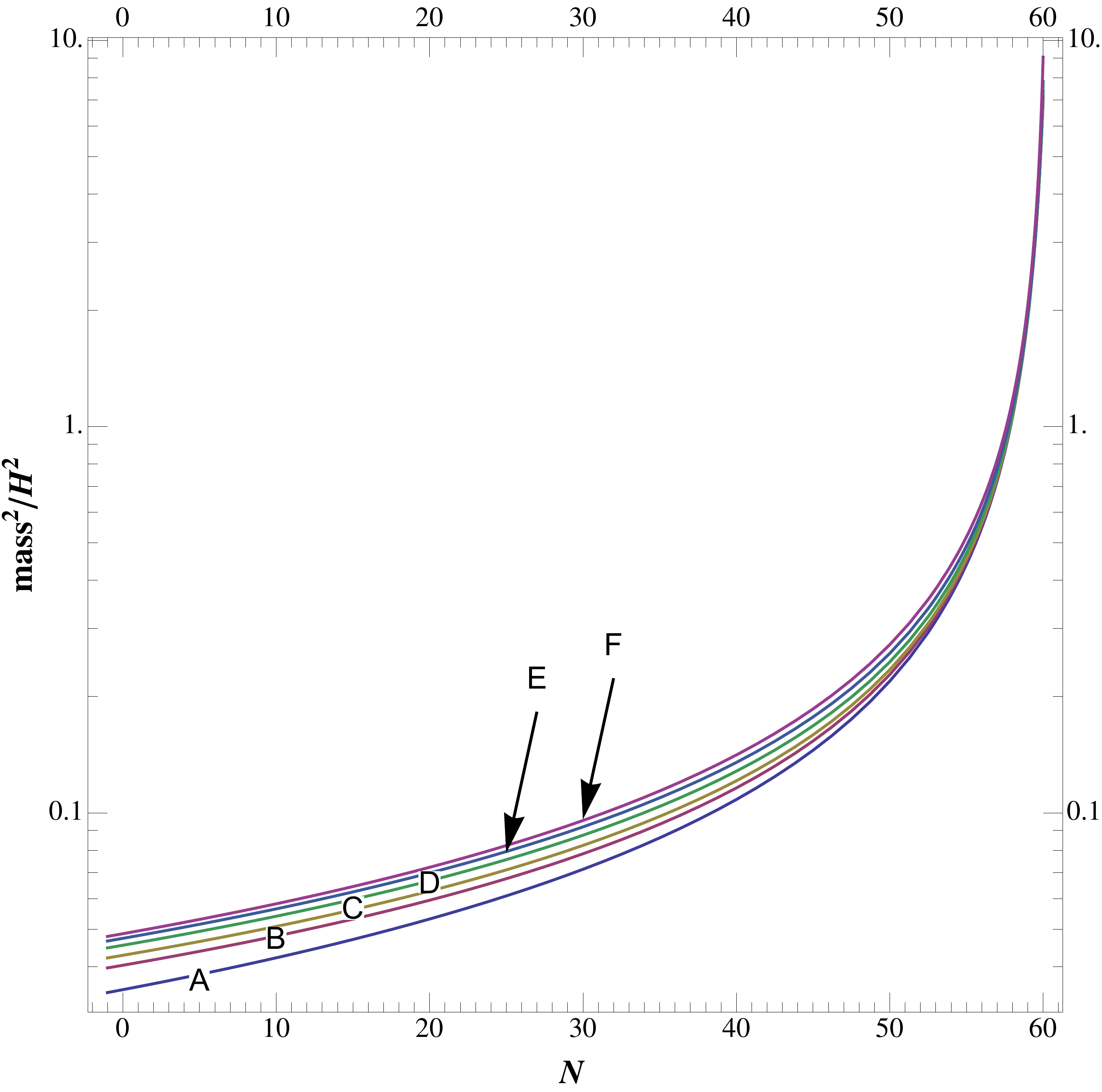}
\end{tabular}
\caption{Left: Time-dependent ratio between the instantaneous power spectra of the isocurvature and curvature perturbations for the mode $k_{60}$ that leaves the Hubble radius 60 e-folds before the end of inflation, for the six trajectories in Fig.~(\ref{f1}). Right: corresponding effective entropic squared masses in Hubble unit (Eq.~\refeq{ms2}).}
\label{adiabatic-limit}
   \end{center}
\end{figure}

As we have stressed in the introduction, a multifield model of inflation is not predictive by itself in general, unless one prescribes a scenario for (p)reheating. \textit{A priori}, there may remain isocurvature perturbations by the end of inflation that can still affect the time evolution of the cosmological fluctuations, and of the curvature perturbation in particular, during reheating or at a later stage. This is what occurs \textit{for example} in modulated (p)reheating \cite{Dvali:2003em,Kofman:2003nx,Bernardeau:2004zz} and its numerous variants, or in the curvaton scenario \cite{Lyth:2001nq}.\\

To check that an adiabatic limit has indeed been reached at the end of inflation in the model studied in the previous section, and to understand its origin, we begin by plotting in Fig.~\ref{adiabatic-limit} (left) the time evolution of the ratio between the instantaneous power spectra of the isocurvature and curvature perturbations for the mode $k_{60}$, for the six trajectories in Fig.~\ref{f1}. For each of them, this ratio decreases from the period of Hubble crossing to the end of inflation, more pronouncedly during the last $5$ e-folds of inflation, to reach a small value of order $10^{-8}$. This suppression of the entropic mode compared to the adiabatic one, also visible in Fig.~\ref{f2}, explains why $\cal P_{\zeta}$ reaches a constant value, and why one was able in the previous section to make predictions for $n_s$ and $r$ for the corresponding models, which was not guaranteed \textit{a priori}. For example, we will exhibit and study in section \ref{sec:oscillations} a two-field inflationary model in which the adiabatic limit is not reached by the end of inflation, and the consequences it has. Note eventually that, strictly speaking, even for the model studied in this section, one should bear in mind that violent processes during (p)reheating, such as parametric resonances, might enhance the entropic fluctuations from their suppressed values at the end of inflation. We will not consider this possibility in the following though.\\
\begin{figure}
  \begin{center}
  \includegraphics*[width=11.5cm]{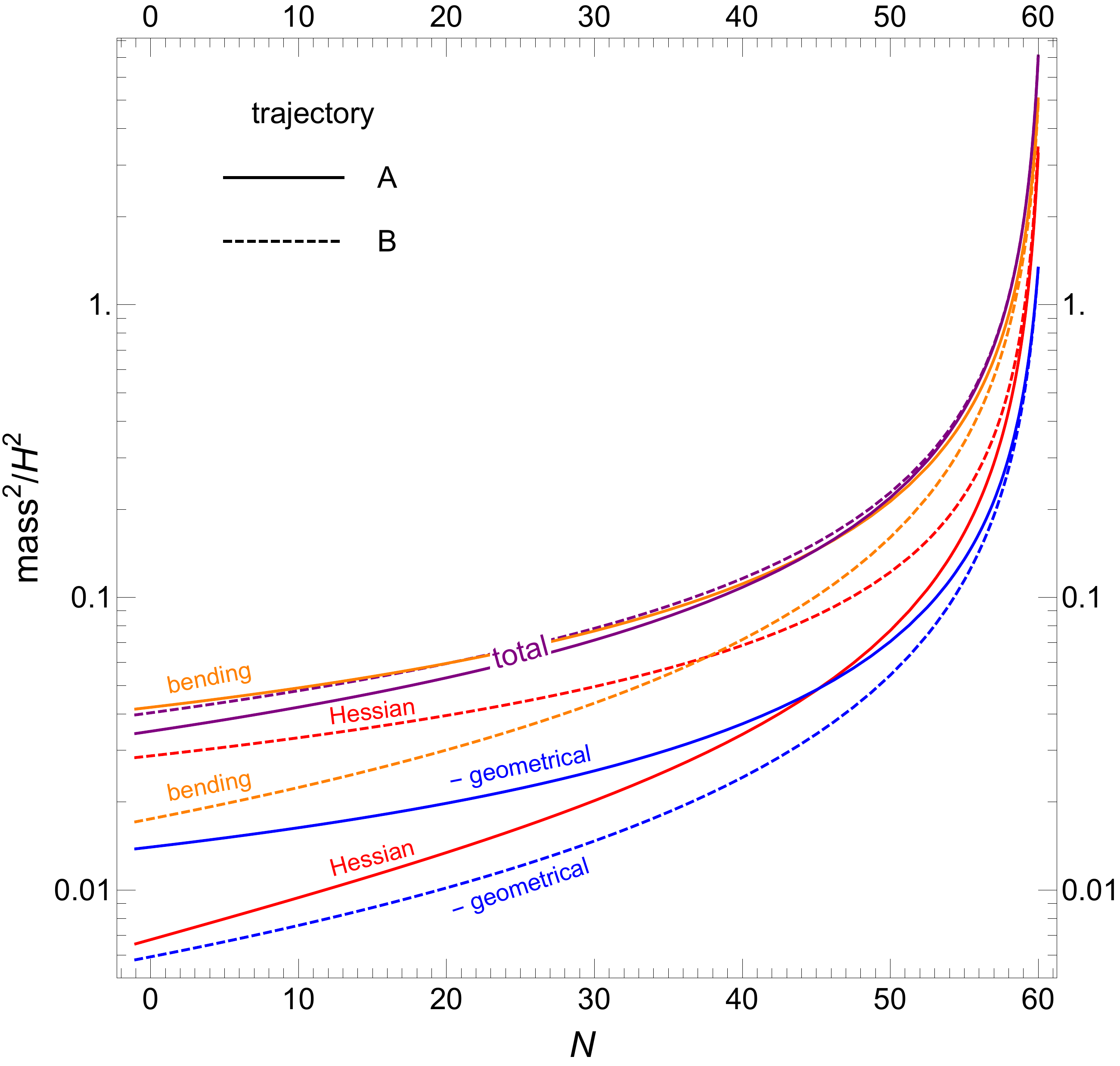}
\caption{Instantaneous Hessian, bending, and geometrical contributions to the super-Hubble effective entropic mass squared in Hubble units $m^{2}_{s {\rm (eff)}}/H^2$ (Eq.~\ref{ms2}), for trajectories {\sf A} and {\sf B}.}
\label{entropicmasses}
    \end{center}
\end{figure}

The super-Hubble evolution of the entropic fluctuation, governed by Eq.~\refeq{eqQs}, is dictated by its effective mass squared Eq.~\refeq{ms2}. By plotting it (in Hubble units) in Fig.~\ref{adiabatic-limit} (right) for the six trajectories {\sf A} to {\sf F}, we witness a similar trend: the entropic direction is light ($m_{s {\rm (eff)}} \ll H$) around Hubble crossing and during the following 30 e-folds; its effective mass grows as inflation proceeds, to reach values of order of the Hubble parameter 5 e-folds before the end inflation; at which point it grows very rapidly, in agreement with the suppression of the entropic mode that we saw on the right panel of Fig.~\ref{adiabatic-limit}. However, this apparent universality does not reflect the variety of situations provided by the six trajectories under study. To understand this, let us study in general terms the three contributions to the effective entropic mass squared in Hubble units $m^{2}_{s {\rm (eff)}}/H^2$ in Eq.~\refeq{ms2}:\\

\noindent \textbullet \,\textbf{Hessian contribution}.---The first contribution --- $V_{;ss}/H^2$ --- is the `naive' expectation for an effective mass: the second derivative of the potential along the corresponding direction. Note that it is not too naive though, in that in properly takes into account the non-trivial field space metric through the use of  covariant derivative with respect to it. More generally, in multiple field inflation with an arbitrary number of fields, the eigenvalues of the Hessian matrix of the potential, properly normalized by the inverse of the field space metric and the Hubble scale --- $G^{IK}V_{;KJ}/H^2$ --- are the first quantities to calculate to get a picture of the multifield phenomenology of the model. In Ref.~\cite{McAllister:2012am} for example, the knowledge of the mass spectrum of a statistical ensemble of six-field inflationary models easily enabled one to identify these models as belonging to the class of quasi-single-field inflation \cite{Chen:2009we,Chen:2009zp}, with all eigenmasses but one of order one. Here, on the contrary, the Hessian contribution to the mass matrix (always in Hubble units) is much smaller than one around the time of Hubble crossing, indicating a standard multifield scenario with this respect. \\

\noindent \textbullet \,\textbf{Bending contribution}.---The contribution $+ 3 \eta_\perp^2$, due to the instantaneous being of the trajectory, is always positive. Typically neglected in slow-roll type analyses, this contribution can be important by definition in models with significant bending of the trajectory. It implies that a turn in field space, necessary to achieve a transfer from entropic to adiabatic fluctuations, also typically tends to decrease the amplitude of the entropic fluctuations, which in turn reduces their potential impact. This competition between two opposite effects has also been noted and studied in Ref.~\cite{Peterson:2010np} for instance.\\

\noindent \textbullet \,\textbf{Geometrical contribution}.---The last contribution --- $\epsilon \,R^{{\rm field \, space}}$ --- has been somewhat ignored in the literature, although it can have a profound impact on cosmological fluctuations. Let us remind the reader that $\epsilon \equiv -\dot H/H^2$ is the standard inflationary parameter whereas $R^{{\rm field \, space}}$ is the Ricci scalar of the field space metric. Barring fine-tuning, we will consider in the following discussion that the latter is of order one (it equals $-4/3$ in the model under study for instance). The positive parameter $\epsilon$ is smaller than one by definition during inflation, and it is actually much smaller than one during the phase of almost de-Sitter inflation that is usually considered and supported by the data. This `geometrical' contribution to the effective entropic mass is thus typically small around the time of Hubble crossing for the relevant cosmological scales. However, at the end of inflation, which is usually defined to be at $\epsilon=1$, this unavoidable contribution is of order one and it is therefore crucial to take it into account. If $R^{{\rm field \, space}}$ is positive, its effect is to suppress the entropic mode. If $R^{{\rm field \, space}}$ is negative though, it tends to destabilise the entropic direction by rendering its effective mass tachyonic, \textit{whatever the curvature of the potential} in that direction. Hence, similarly to the Hessian matrix, the field space Ricci scalar of a two-field inflationary model --- which can be calculated already at the level of the Lagrangian, independently of a specific trajectory --- reveals an important information about whether an adiabatic limit might be reached or not. This discussion straightforwardly generalises to the case of $N$-field inflationary models, in which the geometrical contribution to the squared mass matrix in Hubble units is $-2 \epsilon \, G^{IK}R_{K L M J} e_{\sigma}^L e_{\sigma}^M$ (\textit{c.f.} Eq.~\ref{masssquared})\footnote{Note that for multifield noncanonical Lagrangians of the form $P(X,\phi^I)$, studied in Refs.~\cite{Langlois:2008mn,RenauxPetel:2008gi}, the geometrical contribution is proportional to $\epsilon/P_{,X}$ and the direct link with the `deceleration' parameter $\epsilon$ is lost.}.\\

Despite the apparent similarity between the effective entropic masses of the six trajectories that we saw on the right panel of Fig.~\ref{adiabatic-limit}, their various contributions vary significantly from case to case. We highlight this by displaying them in Fig.~\ref{entropicmasses} for the representative cases of trajectories {\sf A} and {\sf B}. In the former, the Hessian contribution, which is frequently assumed to dominate the total effective mass, is dwarfed in magnitude both by the geometrical contribution (at least during the first 40 e-folds after Hubble crossing of the cosmological modes), and by the bending contribution, which actually dominates. This bending contribution decreases from trajectory {\sf A} to {\sf F}, to vanish in the latter case of a geodesic motion in field space. As one goes from trajectory {\sf A} to {\sf F}, the model also interpolates from large-field to small-field inflation, so that $\epsilon$, and hence the amplitude of the geometrical contribution around the time of Hubble crossing, decreases (as we have stressed, this contribution is equally important for all trajectories at the end of inflation). Note that there are also similarities between the various trajectories. In all cases, the bending of the trajectories increase, and the fields approach a deeper and deeper valley as inflation proceeds, so that the Hessian contributions increases as well: this effect largely explains the origin of the adiabatic limit in our model. \\

Let us now point out a peculiarity: the cautious reader might have noticed from Fig.~\ref{adiabatic-limit} that the smaller the entropic mass, the more efficient the adiabatic limit, \textit{i.e.} the smaller the ratio between the entropic and the adiabatic mode. This somewhat comes at odds with the natural expectation that a light scalar field prevents the reaching of an adiabatic limit. While it is true that a light scalar field \textit{decoupled} from the rest of the inflationary dynamics prevents it, there is no decoupling here: the entropic degree of freedom is constantly coupled to the curvature one by the continuous bending of the trajectory. A smaller entropic mass implies a slower decay of the entropic perturbation on super-Hubble scales, but it also leads to a more efficient feeding of the adiabatic mode. The two effects therefore act in opposite directions and the net effect for the ratio $\mathcal{P}_\mathrm{iso}/\mathcal{P}_\zeta$ is model-dependent. One should also bear in mind that the impact of the entropic mass is not studied here with everything else fixed, as the degree of bending increases significantly from trajectory {\sf F} to {\sf A}.\\

\begin{figure}
 \begin{center}
\begin{tabular}{cc}
\includegraphics*[width=7cm]{fig2105ba.pdf}
&
\includegraphics*[width=7cm]{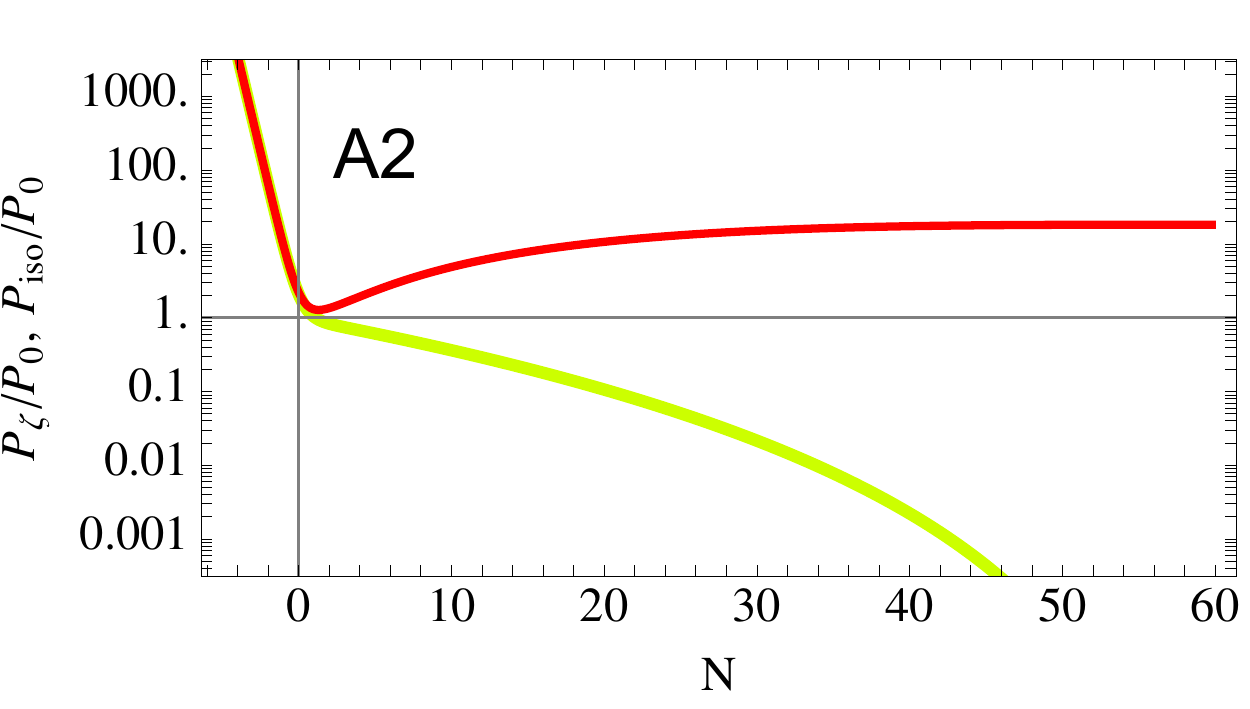}
\\
\includegraphics*[width=7cm]{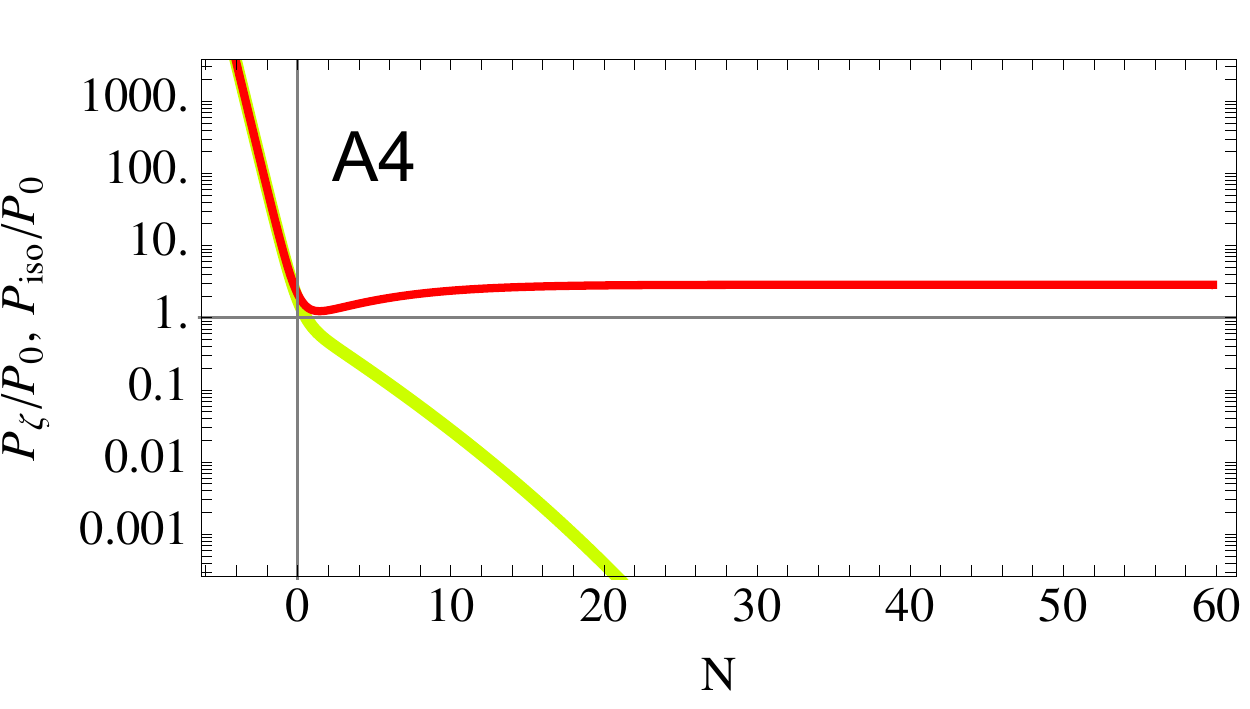}
&
\includegraphics*[width=7cm]{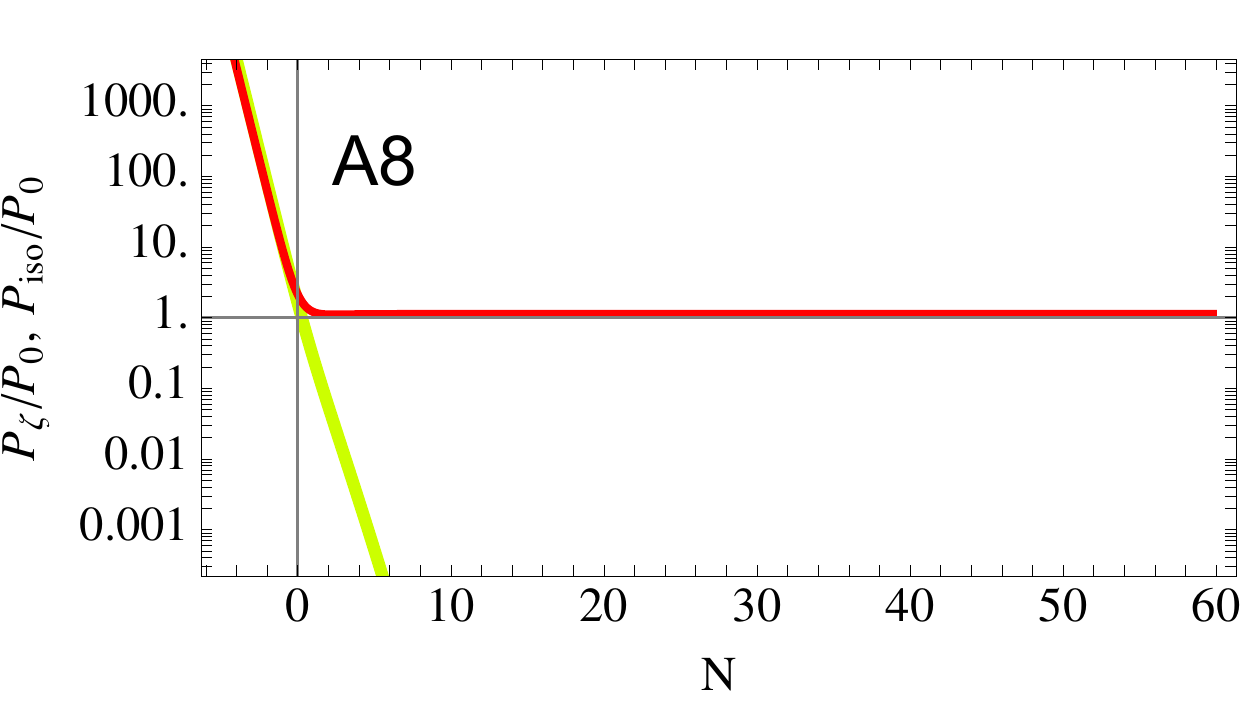}
\end{tabular}
\caption{For trajectories {\sf A}, {\sf A2}, {\sf A4} and {\sf A8}: instantaneous power spectra of the curvature and isocurvature perturbations, $\mathcal{P}_\zeta$ (red lines) 
and $\mathcal{P}_\mathrm{iso}$ (yellow lines), for the mode $k_{60}$. The power spectra are given
in units of the single-field slow-roll prediction \mbox{$\mathcal{P}_0$}  given in Eq.~(\ref{eq:p0}).
 \label{f2mass}}
  \end{center}
\end{figure}

\begin{figure}
\begin{center}
\includegraphics*[width=9cm]{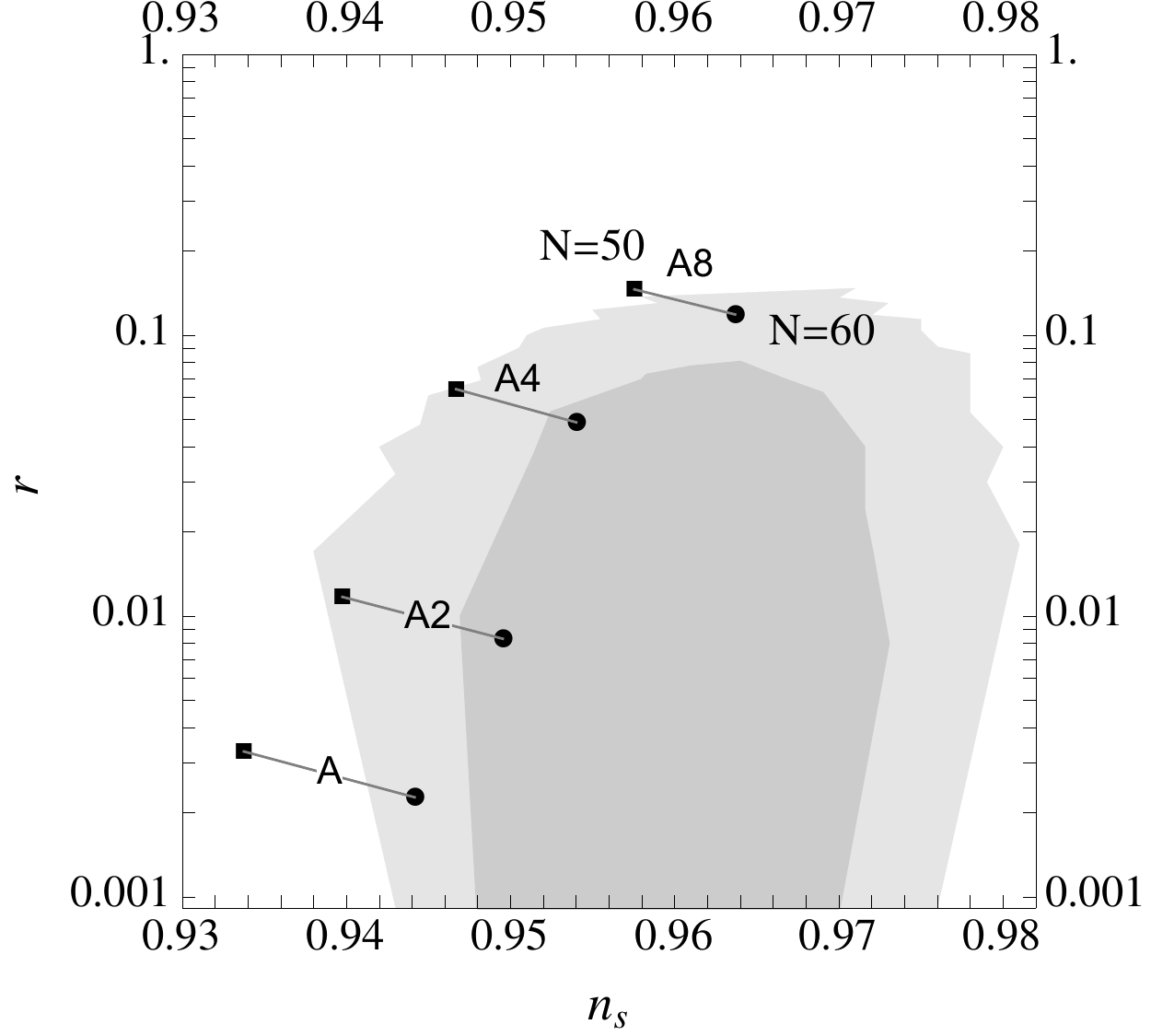}
\caption{Predictions for the scalar spectral index $n_s$ and the tensor-to-scalar ratio $r$ for the 
four trajectories {\sf A} to {\sf A8}, with effectively different masses of the isocurvature
perturbations. The values of $N$ lie between 50~and~60, and the shaded regions correspond to
68\% and 95\% CL constraints from a joint analysis of the Planck and BICEP2 data \cite{Mortonson:2014bja}.
 \label{f3mass}}
 \end{center}
\end{figure}

One way to study the effect of the entropic mass on the cosmological fluctuations is to consider a model with a potential (\ref{eq:pot})
modified by replacing the first mass parameter $m$ by $q\,m$, with $q=2,4,8$, and to study inflationary trajectories starting at $\phi=0$
and giving more than 60 e-folds of inflation (we label them by {\sf A2}, {\sf A4} and {\sf A8}, respectively). Indeed, one can check that all these background are very similar, with an entropic direction mainly along the $\phi$ axis, \textit{i.e.} whose mass is affected by the change $m \to q \,m$. The results for the evolution of the perturbations are given in Fig.~\ref{f2mass} and the predictions for $n_s$ and $r$ are shown in Fig.~\ref{f3mass}. By increasing the mass of the isocurvature fluctuation, its amplitude decays more rapidly, its effect on the curvature perturbation decreases, and the adiabatic limit is reached more rapidly. It is interesting to see in Fig.~\ref{f3mass} how the current experimental sensitivity enables one to directly probe such details of the inflationary Lagrangian.

\section{Impact of spectator fields on cosmological fluctuations}
\label{sec:oscillations}

\begin{figure}
  \begin{center}
  \includegraphics*[width=11.5cm]{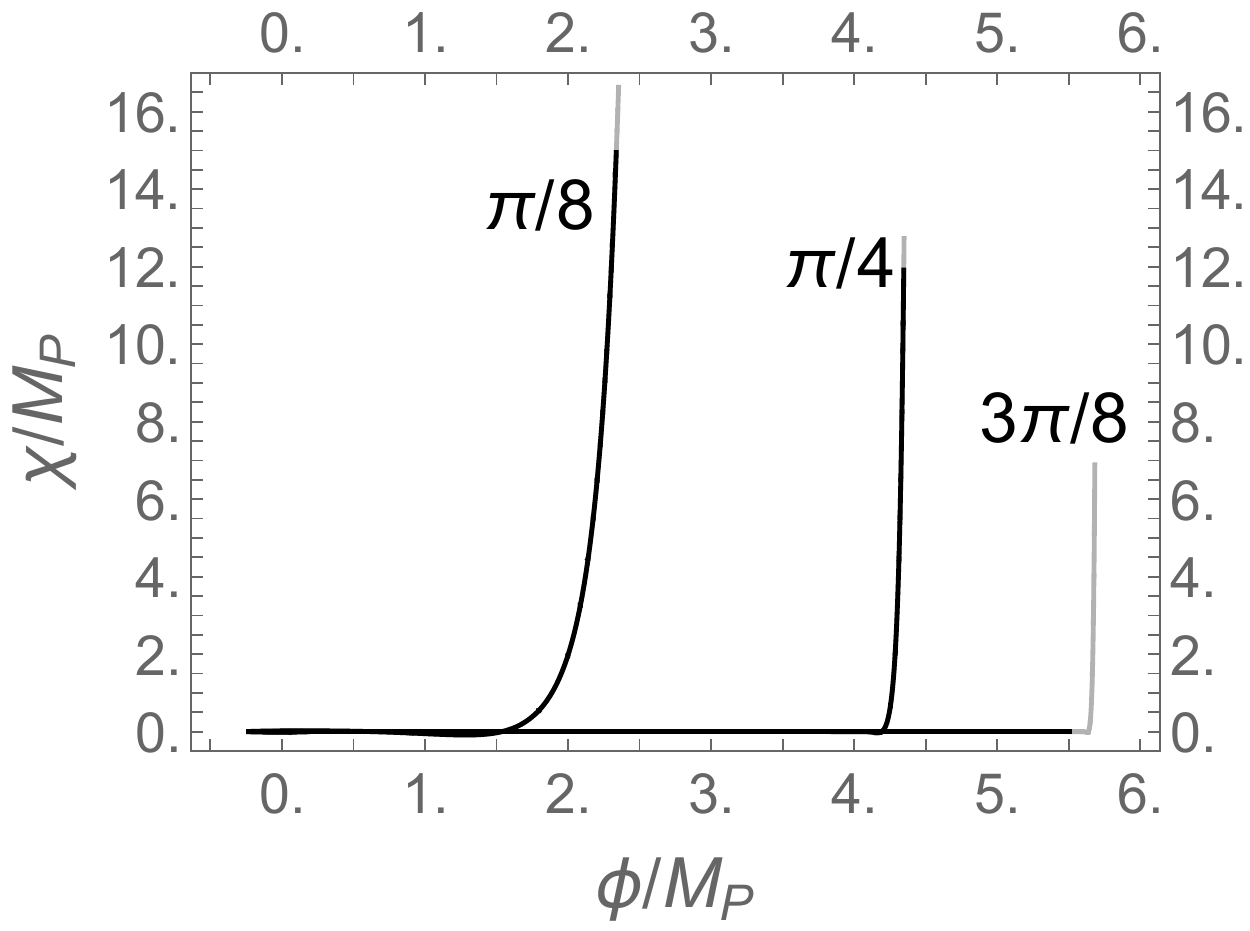}
\caption{A selection of inflationary trajectories for the model (\ref{eq:lgen}) 
with $b(\phi)=0$, with initial conditions $\phi_i=6.15\, {\rm sin}(\theta)$, 
$\chi_i=18\, {\rm cos}(\theta)$, 
and $\theta=\frac{3 \pi}{8}, \frac{\pi}{4}$ and $\frac{\pi}{8}$. Black fragments
correspond to the last 60 e-folds before the end of inflation, characterised by $\epsilon=1$. 
\label{f1aa}}
  \end{center}
\end{figure}

\begin{figure}
\begin{tabular}{cc}
\includegraphics*[width=7cm]{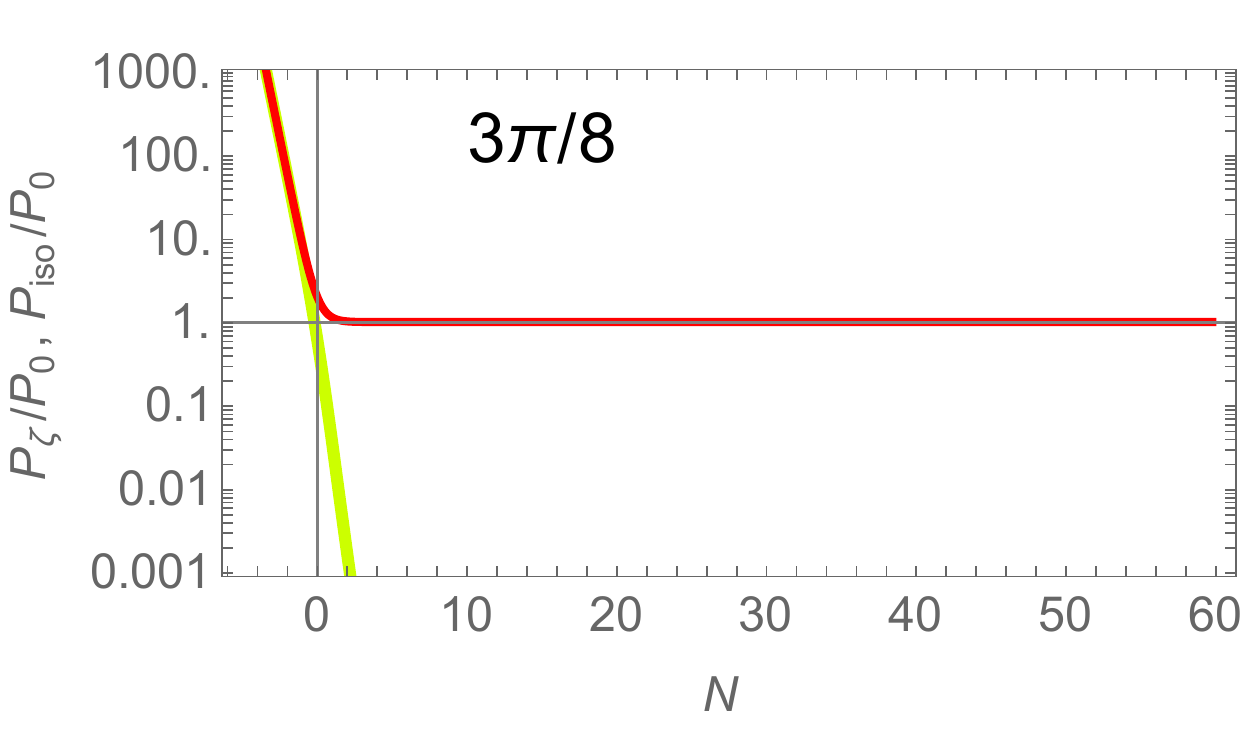}
&
\includegraphics*[width=7cm]{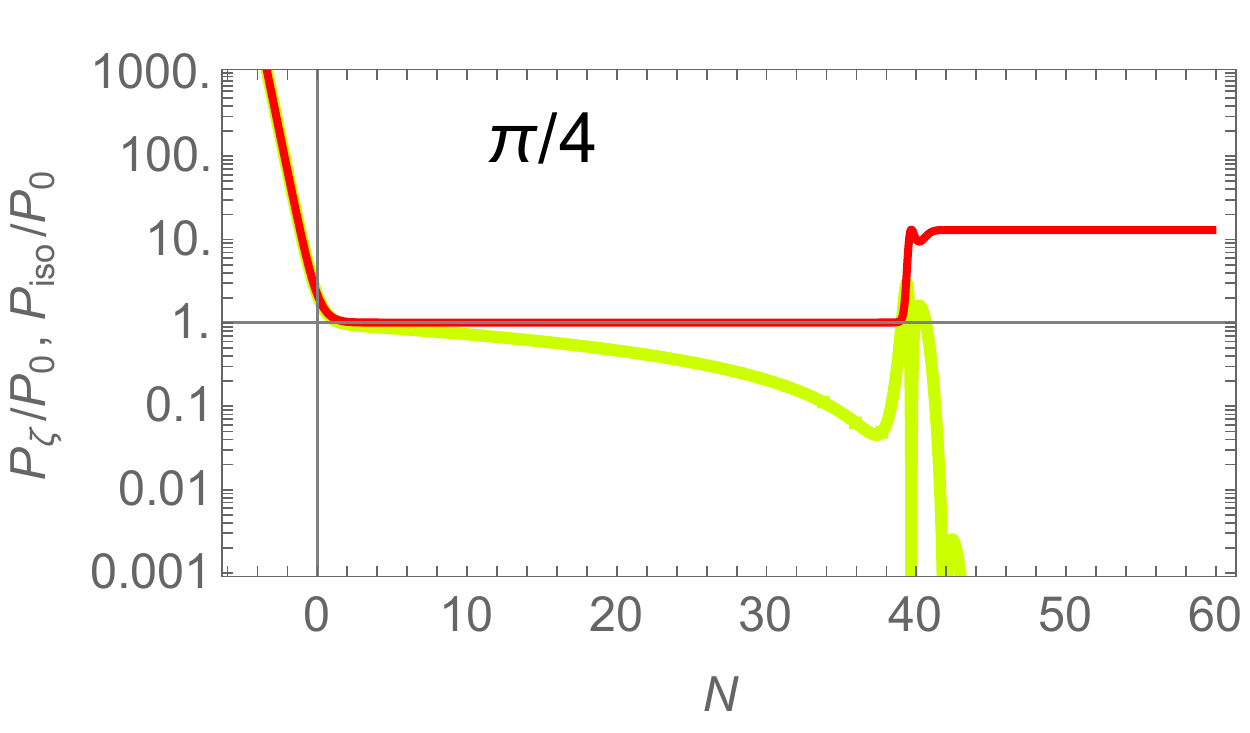}
\\
\includegraphics*[width=7cm]{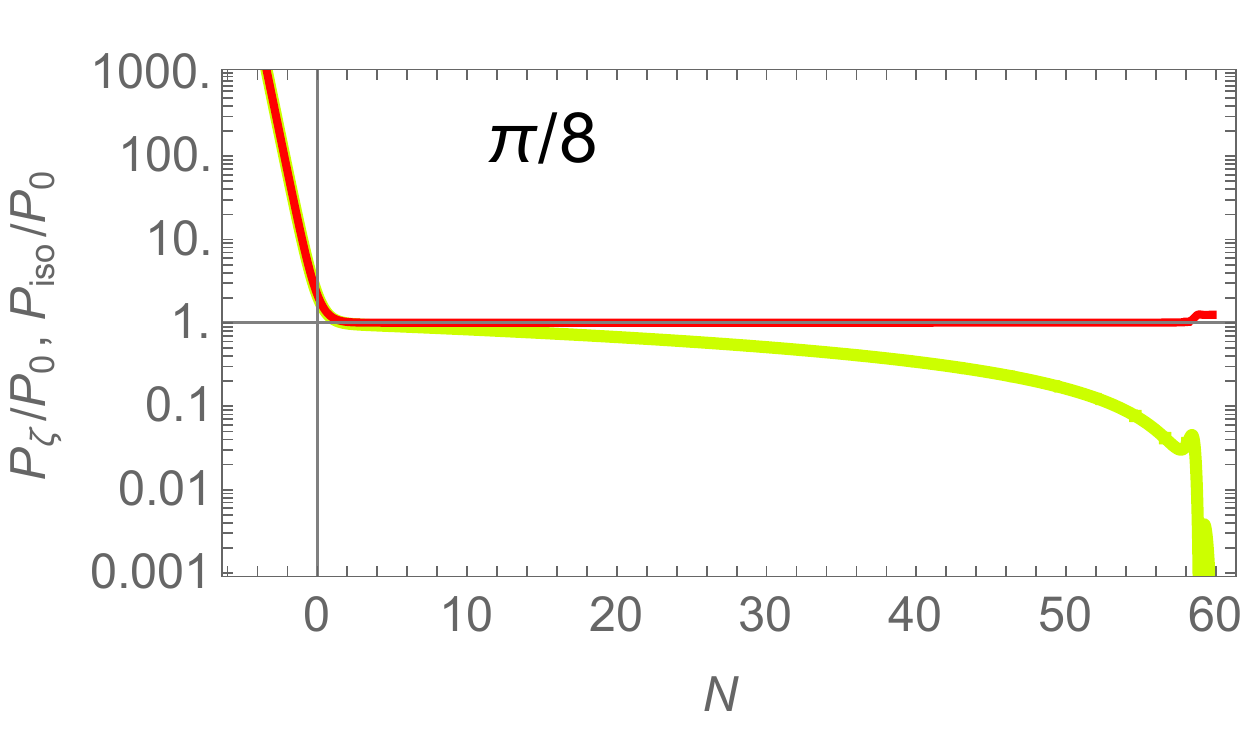}
&
\includegraphics*[width=7cm]{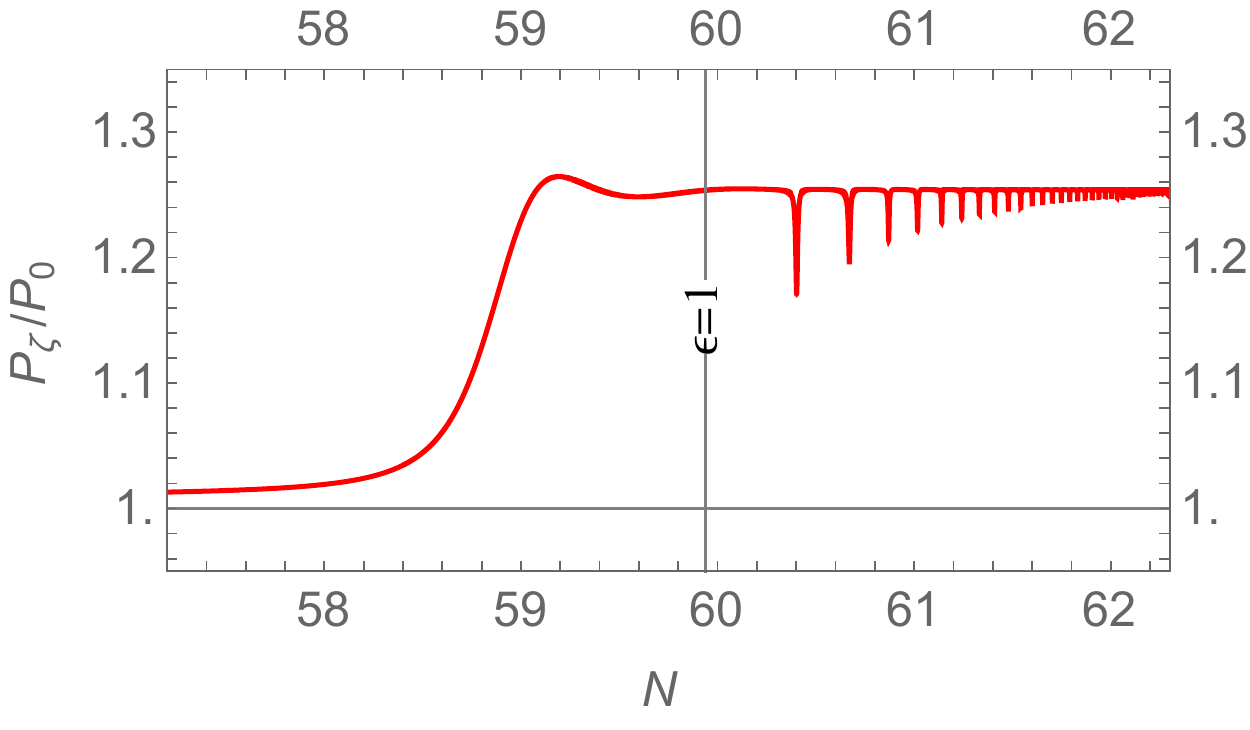}
\end{tabular}
\caption{Instantaneous power spectra of the curvature and isocurvature perturbations, $\mathcal{P}_\zeta$ (red lines) 
and $\mathcal{P}_\mathrm{iso}$ (yellow lines), for the mode $k_{60}$ and the model (\ref{eq:lgen}) 
with $b(\phi)=0$, with initial conditions $\phi_i=6.15\, {\rm sin}(\theta)$, 
$\chi_i=18\, {\rm cos}(\theta)$, 
and $\theta=\frac{3 \pi}{8}, \frac{\pi}{4}$ and $\frac{\pi}{8}$. The last figure zooms in the final period of inflation in the case $\theta=\frac{\pi}{8}$. The power spectra are given
in units of the single-field slow-roll prediction \mbox{$\mathcal{P}_0$}  given in (\ref{eq:p0}). 
 \label{nocoupling}
}
\end{figure}

\subsection{Aspects of multifield phenomenology}

The trajectory {\sf A} in Fig.~\ref{f1} starts at $\phi=0$ with vanishing velocities of the fields. If the field space metric was trivial, these initial conditions would lead to a geodesic trajectory along the $\chi$ direction. The bending contribution to the effective entropic mass, which we saw in Fig.~\ref{entropicmasses} to be dominant for trajectory {\sf A}, would hence vanish. With a trivial field space metric, one may thus expect, more generally, less bending, smaller entropic masses, and therefore a less efficient reaching of the adiabatic limit. For this reason, we consider in this section the same model \refeq{eq:lgen} as previously, but with $b(\phi)=0$, \textit{i.e.} with a trivial field space metric.

We do not carry out an exhaustive study of the initial conditions in this model, but we rather exhibit different types of multifield inflationary phenomenology. In Fig.~\ref{nocoupling}, we plot the instantaneous power spectra of the curvature and isocurvature perturbations, for the mode that leaves the Hubble radius 60 e-folds before the end of inflation, and for the models with initial conditions $\phi_i=6.15\, {\rm sin}(\theta)$, 
$\chi_i=18\, {\rm cos}(\theta)$, 
and $\theta=\frac{3 \pi}{8}, \frac{\pi}{4}$ and $\frac{\pi}{8}$ respectively (see Fig.~\ref{f1aa} for a representation of the respective trajectories).\\

The case $\theta=\frac{3 \pi}{8}$ is typical of a multifield model that is effectively single-field: one can check that the mass of the entropic direction is much heavier than the Hubble scale; the isocurvature fluctuation therefore decays very rapidly, and the curvature power spectrum is unaffected by multifield effects. On the contrary, the trajectory with $\theta=\frac{\pi}{4}$ is typical of the two-field models that can be found in the literature: around $N=40$, a sudden turn in field space leads to the conversion of a light entropic fluctuation to the curvature one, with an increase of its power spectrum by one order of magnitude. Due to the trajectory's change of direction, the entropic direction changes from light to heavy, which explains the rapid decay of the isocurvature perturbation after the turn. This case therefore provides an example of a non-trivial \textit{predictive} multifield model.\\

The third example, corresponding to $\theta=\frac{\pi}{8}$, is particularly interesting. By following the time-evolution of the curvature power spectrum during 55 e-folds after Hubble-crossing of the pivot scale $k_{60}$, one could erroneously conclude that no multifield effect is present in this model. Yet, one can observe a sudden growth of the curvature fluctuation in the last two e-folds of inflation: its power spectrum then grows by $25 \, \%$, within less than one e-fold. While this case might at first sight resemble the one with $\theta=\frac{\pi}{4}$ --- the standard conversion of a light entropic mode to the adiabatic power spectrum through a turn in field space --- it displays a crucial difference: this process arises near the end of inflation, defined as $\epsilon \equiv 1$. As a consequence, one can not expect an adiabatic limit to have been reached by that time. Indeed, evaluating the `final' power spectrum at $N=60$, as is customarily done, is not appropriate here, as the bottom right panel of Fig.~\ref{nocoupling} demonstrates: due to the persistence of an isocurvature mode, the curvature perturbation it still evolving at that time. Solving for the behaviour of the cosmological fluctuations after the end of inflation, one observes that the curvature power spectrum displays a structure of (downwards) spikes of decreasing amplitude. To our knowledge, this type of evolution has not been explored so far and we study it in the rest of this paper.

\subsection{Time oscillatory features of the cosmological power spectra}

\begin{figure}
  \begin{center}
\begin{tabular}{cc}
 \includegraphics*[width=8cm]{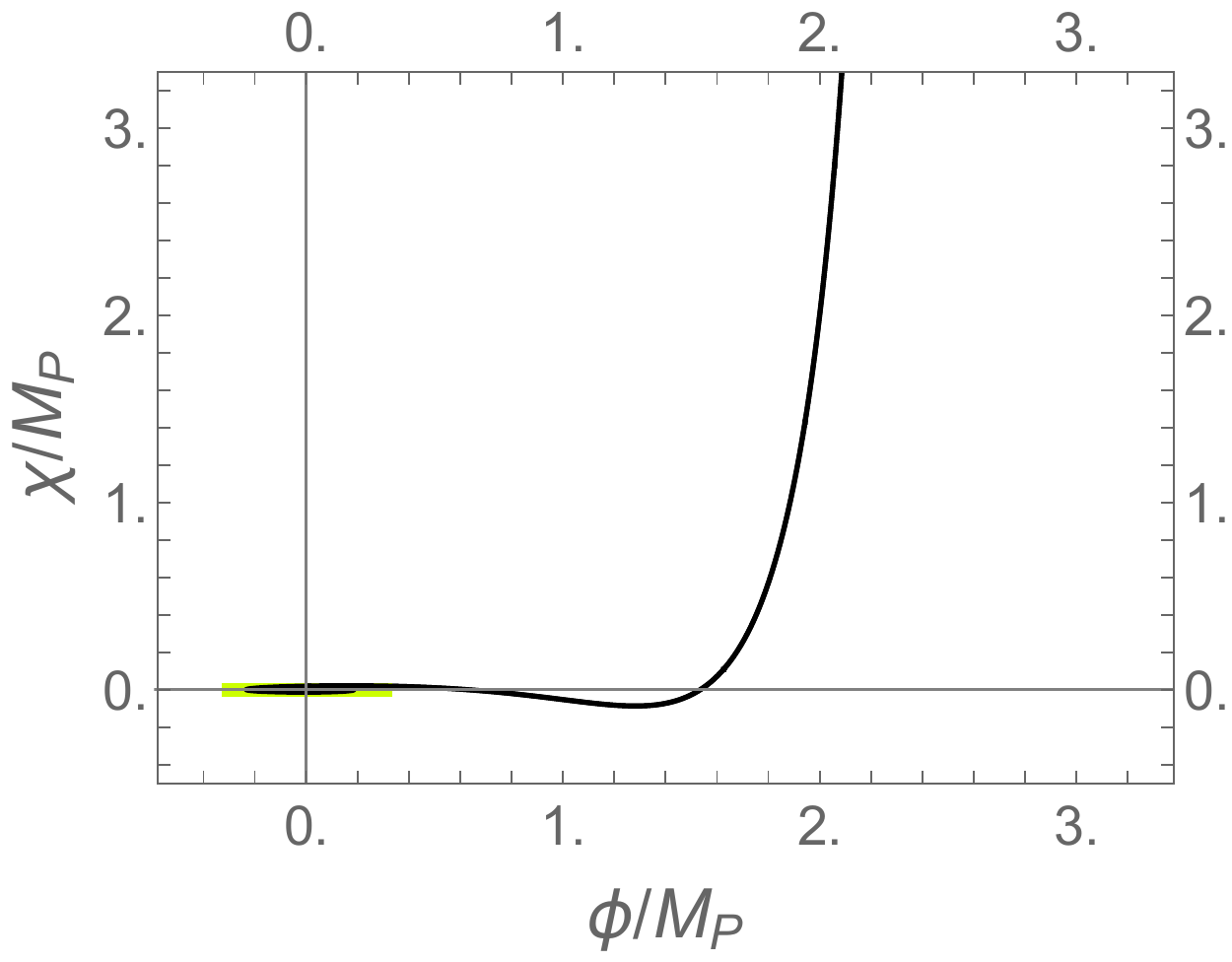}
&
 \includegraphics*[width=8cm]{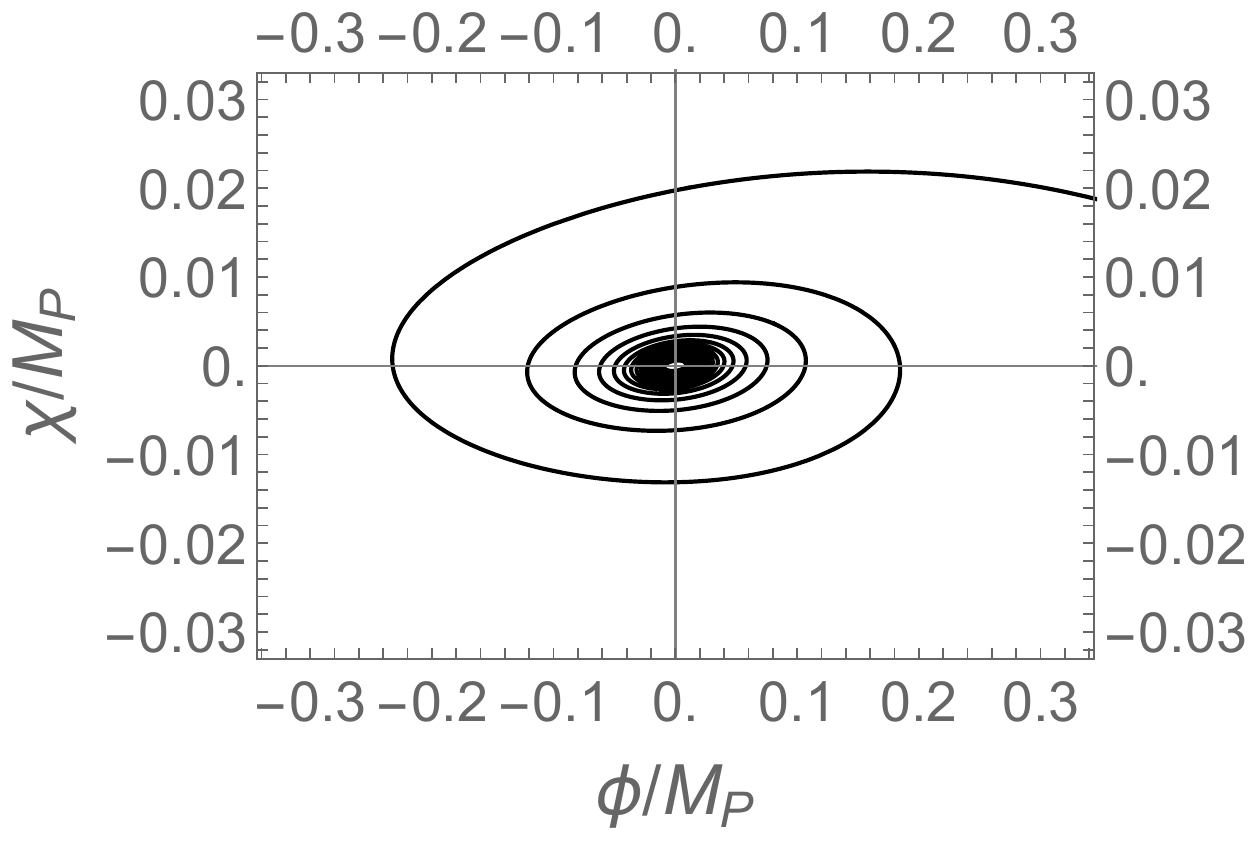}
\end{tabular}
  \caption{Final part of the background trajectory corresponding to $\theta=\pi/8$. The right plot is a blow-up of the shaded area marked in the left plot.}
\label{trajectory-end-inflation}
    \end{center}
\end{figure}

We will actually show that the structure of spikes observed on the bottom right panel of Fig.~\ref{nocoupling} can be very accurately modelled analytically, as well as it can be understood in a much more general context than in the particular model under study. Let us first notice that this phenomenon arises during the phase where the background fields are oscillating around the minimum of the potential at $(\phi,\chi)=(0,0)$ (see Fig.~\ref{trajectory-end-inflation} for a representation of the trajectory between $N=60$ and $N=63$). For such small values of the fields (compared to the relevant Planck scale), it is perfectly valid to perform a local expansion of the potential near the origin. In the following, we thus consider a simple sum of quadratic potentials:
\be
V=\sum_I \half m_I^2 (\phi^I)^2\,.
\label{quadratic-V}
\ee
(with, in our case, $\phi^1 \equiv \phi, \phi^2 \equiv \chi$ and $m_1=m_2=m$). There could be in general cross-terms $m_{IJ}^2 \phi^I \phi^J$ for $I \neq J$ but we do not consider this possibility in the following. Using a standard argument, one can then model the expansion of the universe as if it was matter dominated during the oscillations of the scalar fields (by averaging over many oscillations). In that case, one obtains that the Hubble factor evolves as 
\be
H(t)=\frac{H_i}{1+\frac32 H_i (t-t_i)}\,,
\label{Hana}
\ee
where $t_i$ denotes the initial time at which we apply our modelling, and here and in the following, a subscript $i$ indicates an evaluation at time $t_i$. One can check that for all practical purposes, Eq.~\refeq{Hana} is a very good approximation, with the exact Hubble scale only having small oscillations superimposed onto it. Now, with this Hubble factor, the equations of motion for the scalar fields \refeq{eom-scalars}, which simply read here 
\be
\ddot \phi^I+ 3 H \dot \phi^I+m_I^2 \phi^I=0 \quad ({\rm no\, summation})\,,
\label{background-eoms-quadratic}
\ee
can be exactly solved to find:
\be
\phi^I(t)=\frac{H(t)}{H_i} \left(  \phi^I_i \cos(m_I(t-t_i)) +\frac{1}{m_I}\left( \dot \phi^I_i+\frac32 H_i \phi^I_i \right) \sin(m_I(t-t_i)) \right)\,,
\label{fields-analytical}
\ee
where we imposed the initial conditions. One can verify that Eq.~\refeq{fields-analytical} is an excellent approximation to the exact evolution of the fields, which, in this description, simply evolve as decoupled harmonic oscillators with friction. Let us stress that the local expansion Eq.~\refeq{quadratic-V}, which was crucial to obtain these results, is not very restrictive: it is perfectly valid by definition in small field inflation, whereas for large-field inflation, it usually requires waiting for very few time after the end of inflation, $\epsilon \equiv 1$, at which $\phi^I/M_P ={\cal O}(1)$ (in the case under study for example, the time $t_i$ is taken to be only $0.45$ e-folds after the first reaching of $\epsilon=1$).\\
 
\begin{figure}
 \begin{center}
\begin{tabular}{cc}
\includegraphics*[width=8cm]{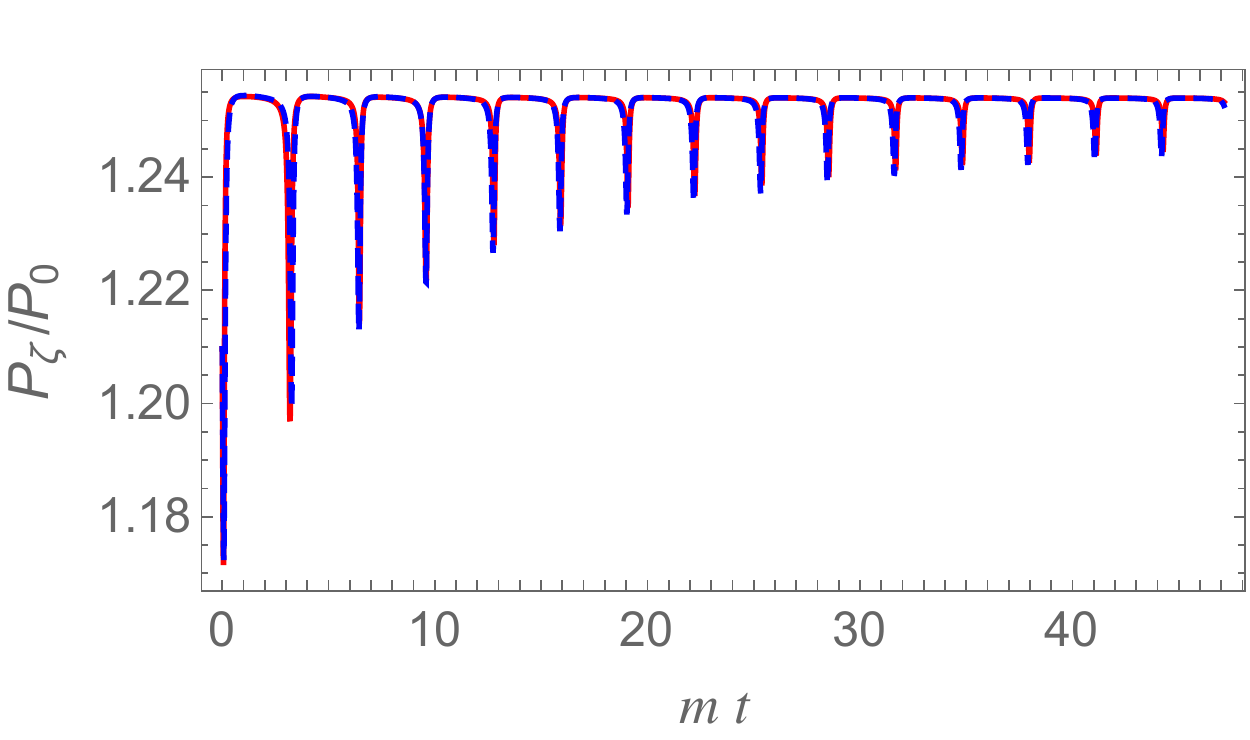}
&
\includegraphics*[width=8cm]{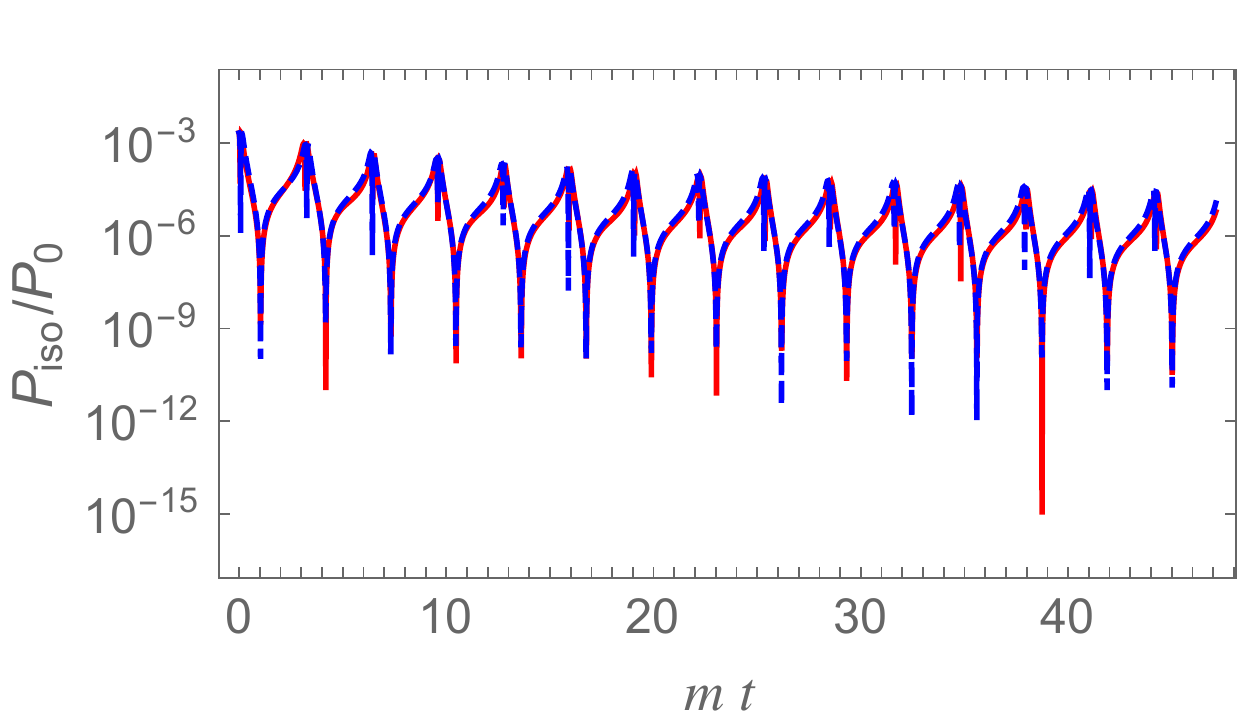}
\end{tabular}
\caption{Left: comparison between the numerically calculated power spectrum of the curvature perturbation (in red) and its semi-analytical counterpart calculated from Eq.~\refeq{constant+decaying} (dashed blue), as a function of time, for the mode $k_{60}$. Right: same comparison for the isocurvature power spectra. We use cosmic time and $t_i=0$ is $0.45$ e-fold after the end of inflation. The agreement is excellent. 
 \label{exact-analytical-spikes}}
  \end{center}
\end{figure}

We now consider perturbations about the above background. For this respect, one can use the separate-universe picture \cite{Sasaki:1995aw,Sasaki:1998ug,Wands:2000dp,Rigopoulos:2003ak}, which states that super-Hubble fluctuations simply behave as perturbations of the background. In this description, the field fluctuations in the spatially flat gauge $Q^I$ thus obey Eq.~\refeq{background-eoms-quadratic}, and hence are given by
\be
Q^I(k,t)=\frac{H(t)}{H_i} \left(  Q^I_i(k) \cos(m_I(t-t_i)) +\frac{1}{m_I}\left( \dot Q^I_i+\frac32 H_i Q^I_i \right) \sin(m_I(t-t_i)) \right)\,.
\label{solutions-QI}
\ee
To be more precise, these are exact solutions of the super-Hubble equations \refeq{pert} in the limit where one can neglect the last term in the mass matrix Eq.~\refeq{masssquared}, \textit{i.e.}, restoring the Planck mass for the sake of the argument, in the limit where one neglect the Planck-suppressed terms in
\be
M^I_{\, J}=m_I^2 \delta^I_{\, J}-\frac{1}{M_P^2} \left(
(3+\epsilon) \dot \phi^I \dot \phi_J+ (\dot \phi^I \ddot \phi_J + \ddot \phi^I \dot \phi_J)/H  \right)\,.
\label{mass-matrix-Planck-suppressed-terms}
\ee
One can check that these terms are indeed negligible for the above background Eq.~\refeq{fields-analytical}, as long as the amplitude of the oscillations are small compared to the Planck mass, which, as we have seen, is a prerequisite of our formalism. However, by neglecting these terms, a crucial property of the cosmological fluctuations is lost: we know that a constant curvature perturbation $\zeta$ and vanishing entropic fluctuation $Q_s$ provides an exact solution to the super-Hubble cosmological dynamics. This indeed trivially solves the relevant equations of motion \refeq{Rdot}-\refeq{eqQs} (see Refs.~\cite{Langlois:2006vv,RenauxPetel:2008gi,Lehners:2009ja} for non-linear extensions of these equations and Refs.~\cite{Wands:2000dp,Rigopoulos:2003ak,Langlois:2005ii,Langlois:2005qp} for non-perturbative statements). On the contrary, the solution Eq.~\refeq{solutions-QI} only describes (oscillating) decaying perturbations, and hence a decaying curvature perturbation $\zeta=\frac{Q_\sigma}{\sqrt{2 \epsilon}}$ ($\epsilon$ oscillates between $0$ and $3$ during the oscillatory phase so the factor $\frac1{\sqrt{2 \epsilon}}$ does not change the argument). Hence Eq.~\refeq{solutions-QI} can not possibly describe an adiabatic limit, and in particular the one that we see arising in the bottom right panel of Fig.~\ref{nocoupling}. What happens here is that, by neglecting the Planck-suppressed terms in the mass-matrix \refeq{mass-matrix-Planck-suppressed-terms}, one does not describe the true cosmological fluctuations, but only their departure from the adiabatic limit solution. As a consequence, this constant mode should be added by hand to Eq.~\refeq{solutions-QI}, as well as the corresponding initial conditions should be modified accordingly: 
\be
Q^I(k,t)=e_\sigma^I(t) \sqrt{2 \epsilon(t)} \zeta_f + Q^I(k,t)^{{\rm decaying}}\,,
\label{constant+decaying}
\ee
where $\zeta_f$ should be read off the numerical simulation, and the $Q^I(k,t)^{{\rm decaying}}$ obey Eq.~\refeq{solutions-QI}, with modified initial conditions such that
\bea
\left(Q_{\sigma} \right)^{{\rm decaying}}_i &=& \left(Q_{\sigma} \right)_i-\sqrt{2 \epsilon_i} \zeta_f \\
\left( \dot Q_{\sigma} \right)^{{\rm decaying}}_i &=& \left( \dot Q_{\sigma} \right)_i-\frac{\dot \epsilon_i}{\sqrt{2 \epsilon_i}} \zeta_f \\
\left(Q_{s n} \right)^{{\rm decaying}}_i &=& \left(Q_{s n} \right)_i \\
\left( \dot Q_{s n} \right)^{{\rm decaying}}_i &=& \left( \dot Q_{s n} \right)_i \,, 
\eea
where $Q_{s n}$ denote the various entropic fluctuation. In our case of interest, we compare in Fig.~\ref{exact-analytical-spikes} the numerically calculated power spectra of the curvature and isocurvature perturbations with their semi-analytical counterparts deduced from Eq.~\refeq{constant+decaying}, for the mode $k_{60}$. We use cosmic time, which is more appropriate than the number of e-folds during this oscillatory phase, and $t_i=0$ is $0.45$ e-fold after the end of inflation. The agreement is excellent, with the two curves being almost indistinguishable, for both types of perturbations. To help visualisation, we stop at $t_f=15 \pi/m$ but the agreement is equally good after that point. Of course, in accord with our derivation of this effect, one can check that the successive minima of the evolving curvature power spectrum coincide with the peaks of $\eta_\perp$, \textit{i.e.} they arise at each turn in field space every $\Delta t=\pi/m$. Like for the estimate of the spectral index in Eq.~\refeq{semi-analytical}, we call our description semi-analytical as it requires the knowledge of the final value of the curvature perturbation $\zeta_f$, that we calculate numerically. We therefore do not predict entirely the behaviour of cosmological fluctuations. However, acknowledging that an adiabatic limit is reached, we are able to model the transient oscillatory structure of the curvature and isocurvature power spectra.

\begin{figure}
 \begin{center}
\begin{tabular}{cc}
\includegraphics*[width=8cm]{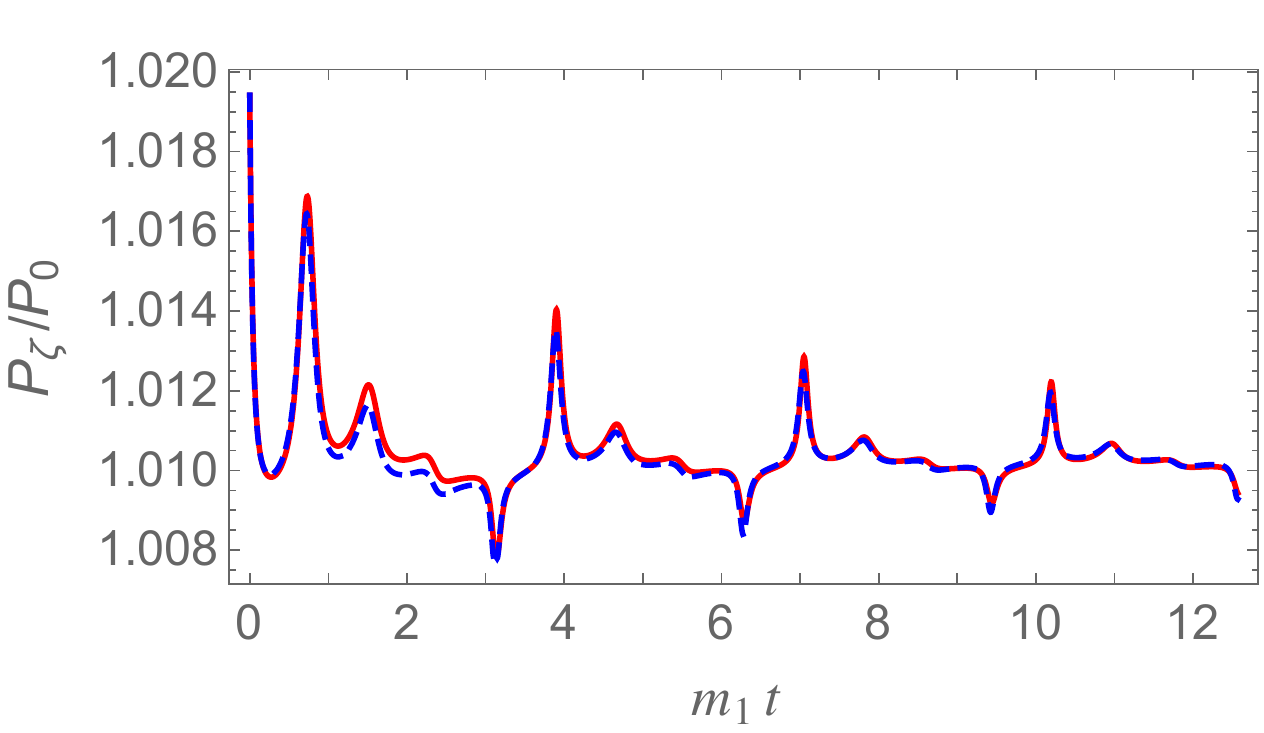}
&
\includegraphics*[width=8cm]{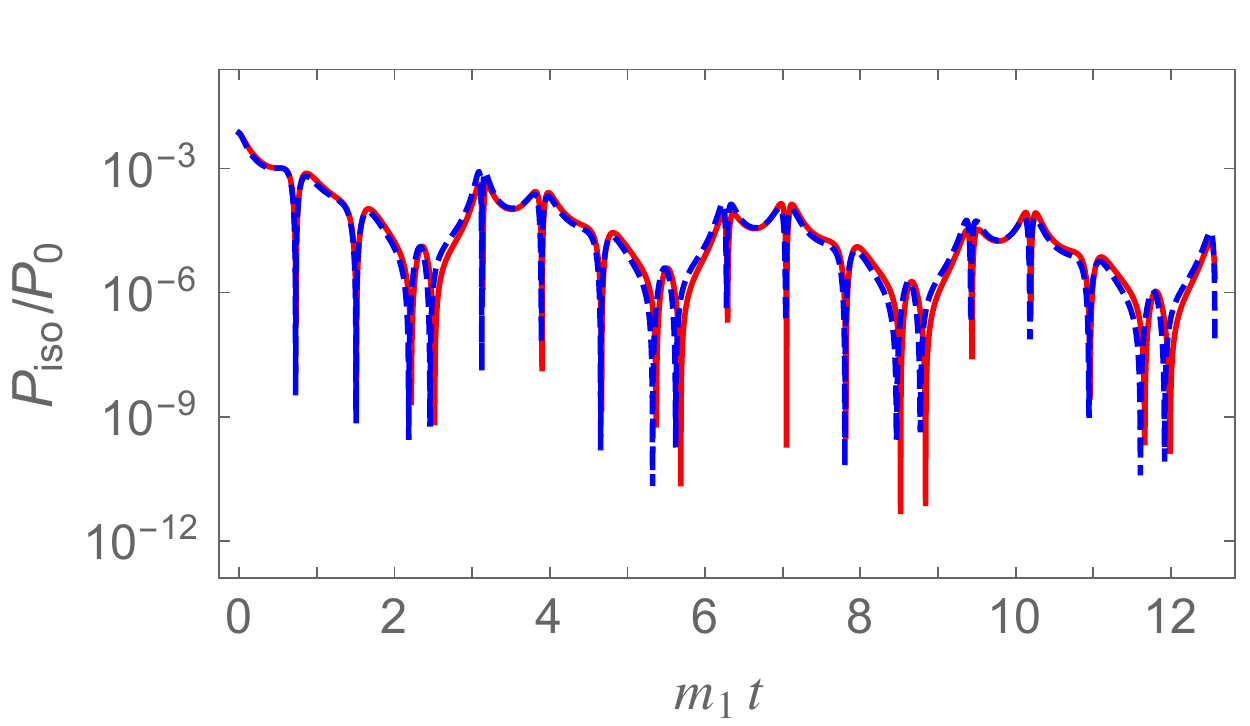}
\end{tabular}
\caption{Same plots as in Fig.~\ref{exact-analytical-spikes}, for the model \refeq{double-quadratic} with initial conditions $\phi_{1,i}=18 \cos(\theta)$, $\phi_{2,i}=18 \sin(\theta)$ and $\theta=4.8/10 \pi$. We use cosmic time and $t_i=0$ is $0.8$ e-fold after the end of inflation. The agreement is very good.
 \label{exact-analytical-spikes-double-quadratic}}
  \end{center}
\end{figure}

As should be clear from our general discussions, we emphasise that the oscillatory structures we have described, and our modelling of them, are not restricted to the particular model that we have considered so far. For example, they arise in the arguably simplest model of inflation with multiple fields, \textit{i.e.} the well-studied model of double quadratic inflation \cite{Polarski:1994rz,GarciaBellido:1995qq,Langlois:1999dw,Rigopoulos:2005ae,Rigopoulos:2005us,Vernizzi:2006ve}, albeit with particular initial conditions. We demonstrate this by considering the inflationary model
\begin{equation}
\mathcal{L} = -\frac{1}{2}\partial_\mu\phi_1\partial^\mu\phi_1-\frac{1}{2}
\partial_\mu\phi_2 \partial^\mu\phi_2 - \frac12 m_1^2 \phi_1^2 - \frac12 m_2^2 \phi_2^2
\label{double-quadratic}
\end{equation}
with $m_2=4 m_1$, and initial conditions $\phi_{1,i}=18 \cos(\theta)$, $\phi_{2,i}=18 \sin(\theta)$ and $\theta=4.8/10 \pi$ (and vanishing velocities like before). These parameters and initial conditions are chosen in such a way that the heavy field dominance ends just before the end of inflation: contrary to the bulk of initial conditions for this model, for which inflation initially proceeds along the heaviest direction ($\phi_2$), and then along the lightest one ($\phi_1$), the initial small value of the the latter field is such that it can not sustain a period of inflation. After the phase of heavy field dominance, the two fields therefore oscillate around the bottom of the potential, as described by Eq.~\refeq{fields-analytical}. It is therefore not a surprise that the same type of oscillatory patterns as before arise in the time-dependent curvature and isocurvature power spectra. We represent them in Fig.~\ref{exact-analytical-spikes-double-quadratic} (in red), together with their semi-analytical counterparts deduced from Eq.~\refeq{constant+decaying} (in blue), for the mode $k_{60}$, and with initial time $t_i$ $0.8$ e-folds after the end of inflation: the agreement is again very good. Two types of oscillatory structures, of frequencies $\omega_1=m_1$ and $\omega_2= m_2= 4 \omega_1$, are clearly visible. It is interesting that the resulting typical beat pattern enables one to directly infer from these plots the ratio between the two masses $m_2/m_1=4$.\\

As observations ultimately constrain the properties of the primordial fluctuations deep in the radiation era, one could wonder why our modeling of their transient behaviours at the end of inflation is useful at all. We find it relevant for the following reasons:
\begin{figure}[!h]
\begin{center}
\begin{tabular}{cc}
\includegraphics*[width=7cm]{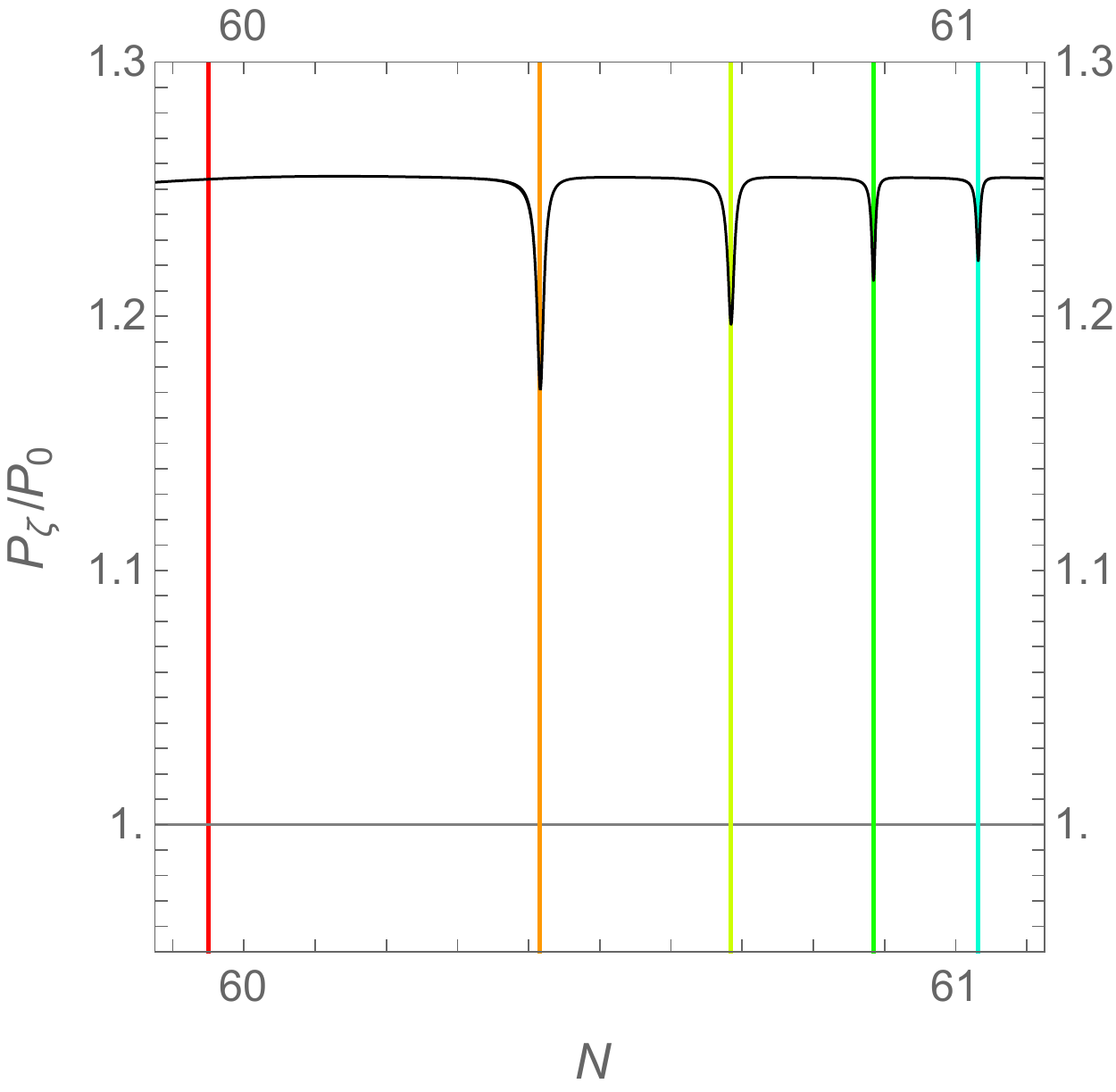}
&
\includegraphics*[width=7cm]{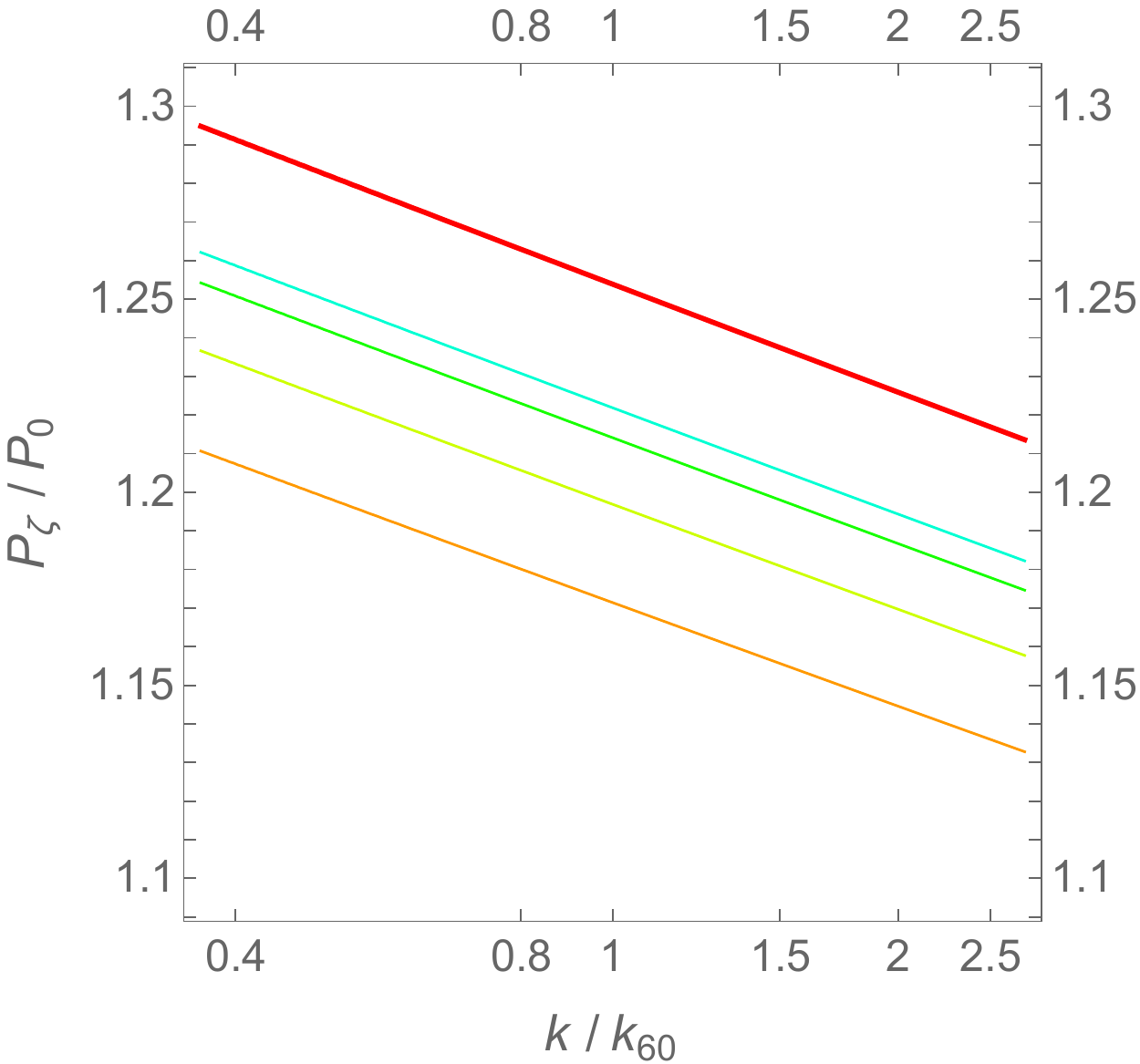}
\end{tabular}
\caption{Influence of the time of evaluation on the properties of the cosmological fluctuations. Left: time-dependent curvature power-spectrum for the mode $k_{60}$. The coloured vertical lines represent various times of evaluation for the calculation of the curvature power-spectrum as a function of scale, represented on the right panel with the same colour coding.}
\label{Dependence-time-evaluation}
   \end{center}
\end{figure}
it is often assumed, for simplicity, that the phase of reheating is sufficiently rapid that thermal equilibrium is reached immediately after the end of inflation (like in instant preheating \cite{Felder:1998vq} or in two-field hybrid inflation \cite{Alabidi:2006wa,Sasaki:2008uc}), or that between the end of inflation and the onset of the radiation-dominated era the curvature perturbation remains constant on super-Hubble scales and the energy density of the universe is dominated by that of the oscillating inflaton field (which eventually decays to Standard Model particles). In these descriptions, one simply identifies the curvature power spectrum at the end of inflation with the observable one, \textit{i.e.} the one relevant for setting initial conditions at the beginning of the radiation era. However, choosing to evaluate the perturbations precisely at the end of inflation, at the first reaching of $\epsilon=1$, or at a slightly later time, may entail different predictions. We demonstrate this arbitrariness in Fig.~\ref{Dependence-time-evaluation}, where we represent, for the first model considered in this section, the curvature power spectrum as a function of scale (right), evaluated at the various times indicated on the right by the colored vertical lines ($N=60$ still represents the end of inflation). Note that, independently of the time of evaluation, the time-dependent oscillatory features of the fluctuations do not lead to an oscillatory power spectrum as a function of scale: the various power spectra depicted on the right panel of Fig.~\ref{Dependence-time-evaluation} can be very accurately described by a standard power law. However, the overall amplitude of the power spectrum obviously depends on the time of evaluation. This exemplifies that extracting the parameters of the inflationary Lagrangian, for example the mass scale $m$ in our case, can not be achieved within the approximation of the instantaneous passage between inflation and the reheating period if the adiabatic limit has not been reached by the end of inflation. Moreover, as should be clear from the semi-analytical estimate \refeq{semi-analytical}, not only does this arbitrariness affect the overall amplitude of the curvature power spectrum, but it also alters the corresponding spectral index. This effect is small here: one finds $n_s=0.9676$ (red), $n_s=0.9668$ (orange), $n_s=0.9670$ (yellow), $n_s=0.9672$ (green) and $n_s=0.9673$ (blue). It is nonetheless present in general and should be taken into account. Let us eventually note that the purely multi-field effects discussed here are convoluted with other theoretical uncertainties due to the modeling of the reheating phase; the latter arise in generic models of inflation, even single-field, for example because of the need to accurately identify the moment of Hubble crossing of the relevant cosmological scales, see \textit{e.g.} Refs.~\cite{Martin:2006rs,Martin:2010kz,Adshead:2010mc,Easther:2011yq,Martin:2014nya}.

\section{Conclusions}
\label{sec:conclusion}

Efforts to embed the inflationary paradigm into ultraviolet-complete theories require considering inflationary scenarios with multiple degrees of freedom (see \textit{e.g.} Ref.~\cite{Baumann:2014nda}). It is then necessary to properly take into account the isocurvature fluctuations that arise in addition to the adiabatic curvature perturbation, and their couplings to the latter. A failure to do so might result in completely wrong observational predictions, as we have demonstrated in a particular but representative example of a two-field inflationary model in no-scale supergravity, concentrating on the scalar spectral index and the tensor-to-scalar ratio. We have shown that the current observational sensitivity on these quantities is such that a slight alteration of the Lagrangian parameters, such as a multiplication of a mass parameter by two, might change a model from being excluded to being favoured by data.\\

We have also stressed that multifield inflationary scenarios should \textit{a priori} be considered as non-predictive, unless one supplies a prescription for the post-inflationary era, or one establishes that an adiabatic limit, in which isocurvature perturbations have decayed, is reached before the end of inflation. We have studied the various contributions to the effective entropic masses squared in Hubble units, which governs the behaviour of the entropic modes on super-Hubble scales. Besides a standard Hessian contribution, and a bending contribution --- due to the deviation of the trajectory from a geodesic in field space --- we have emphasised the importance of a so-called geometrical contribution. The latter is proportional to the `deceleration' parameter $\epsilon$ and, in two-field models, to the Ricci scalar of the field space metric $R^{{\rm field \, space}}$. Barring fine-tuned models, this contribution is therefore of order one at the end of inflation. Depending on the sign of $R^{{\rm field \, space}}$, it hence can act in the direction of rendering the isocurvature direction more massive, or of destabilising it. We emphasise that this important piece of information as far as the adiabatic limit is concerned --- and therefore the predictability of the model at hand --- can be directly extracted at the level of the inflationary Lagrangian, independently of a specific trajectory.\\

In the last section, we have studied the impact on cosmological fluctuations of a light spectator field: a field lighter than the Hubble scale around Hubble crossing, which does not participate in the background dynamics during the bulk of inflation, but that starts oscillating around the bottom of its potential at the end of the former. We have shown that it leads to decaying oscillatory features in the time-dependent curvature and isocurvature power spectra, that we were able to model semi-analytically in a very accurate manner. We are well aware that the simple models that we studied lack a description of the phase of (p)reheating. We did not take into account the coupling of our inflaton fields, neither to themselves nor to other degrees of freedom. Studying the corresponding processes of particle production was beyond the scope of this work, but we think that our modeling is anyway relevant, for at least two reasons: it is necessary to take it into account if the phase of reheating is delayed after a few oscillations of the background fields, which can arise when the couplings of the inflaton fields to the other degrees of freedom (or self couplings) are sufficiently weak. It also demonstrates that the often-used approximation of an abrupt end of inflation, together with the arbitrariness at which one then chooses to evaluate the cosmological power spectra, entails important theoretical uncertainties on the observational predictions, such as on the overall amplitude of the primordial fluctuations and on the scalar spectral index. It remains to be seen whether the oscillatory features that we explained might lead, in different contexts, to even more dramatic observational consequences, such as features in the $k$-dependent power spectra. We leave this open question for further study.

\paragraph{Acknowledgements.} KT acknowledges useful correspondence with J.~Ellis. We thank D.~Langlois for useful discussions. This work was partly supported by the grant  IP2011 056971 
from the Ministry of Science and Higher Education of Poland, and by French state funds managed by the ANR within the Investissements d'Avenir programme under reference ANR-11-IDEX-0004-02.

\end{document}